\documentclass{article}

\usepackage{PRIMEarxiv}
\usepackage{epstopdf}
\usepackage[caption=false]{subfig}
\usepackage[utf8]{inputenc} 
\usepackage[T1]{fontenc}    
\usepackage[hidelinks]{hyperref}       
\usepackage{url}            
\usepackage{booktabs}       
\usepackage{amsfonts}       
\usepackage{nicefrac}       
\usepackage{microtype}      
\usepackage{amssymb}        
\usepackage{lipsum}
\usepackage{fancyhdr}       
\usepackage{graphicx}       
\usepackage{subcaption}
\usepackage{tabularx}
\usepackage[numbers,sort&compress]{natbib}
\usepackage{amsmath,bm}

\title{Assessment of tabulated-chemistry models for lean premixed strained hydrogen flames with low-dimensional manifolds}

\author{
  A. Porcarelli$^{a,*}$, P.E. Lapenna$^{b}$, F. Creta$^{b}$ and I. Langella$^{a}$ \\
  $^{a}$Faculty of Aerospace Engineering, TU Delft, Kluyverweg 1, 2629 HS, Delft, Netherlands \\
  $^{b}$Department of Mechanical and Aerospace Engineering, Sapienza, University of Rome, \\ Via Eudossiana 18, 00189, Rome, Italy \\
  \texttt{*\href{mailto:a.porcarelli@tudelft.nl}{a.porcarelli@tudelft.nl}} \\
}

\begin{document}
\maketitle

\begin{abstract}
This study presents a comprehensive \textit{a priori} analysis of tabulated-chemistry models for both laminar and turbulent lean premixed hydrogen flames in strained counterflow configuration. Particular focus is drawn on differential and preferential diffusion effects and the synergistic interaction of thermodiffusive instabilities and turbulence that existing models struggle to capture. Through detailed assessment of various modelling approaches at unfiltered and filtered grids, we identify significant limitations in traditional unstretched flamelet manifolds, particularly their strong filter dependence and systematic reaction rate mispredictions. To address these challenges, we introduce and evaluate novel strained flamelet approaches, including: (1) a one-dimensional manifold constructed from a single strained flamelet that provides computationally efficient and reliable consumption speed predictions at coarser grids, and (2) a two-dimensional manifold combining fixed strain with varying equivalence ratio that demonstrates improved performance in predicting the local reaction rates across multiple grid resolutions. Additionally, we develop a correction methodology derived from laminar simulations that significantly improves consumption speed predictions of unstretched flamelet manifolds in turbulent settings. Unlike previous works, our solutions maintain computational efficiency without increasing manifold dimensionality, keeping memory costs unchanged. These advancements provide guidance for developing reliable LES models that properly account for differential and preferential diffusion and strain effects in practical hydrogen combustion systems.
\end{abstract}

\begin{keywords}
hydrogen; strained flames; tabulated-chemistry; presumed filtered density function (FDF); F-TACLES
\end{keywords}

\section{Introduction} \label{sec:intro}
Hydrogen technologies have gained significant momentum recently, offering a carbon-free solution for sectors where electrification is challenging, such as aviation and other transportation industries~\cite{IEA2024world,IEA2024global}. Within the realm of hydrogen applications, recent scientific advancements focus on fully~\cite{reichel2015investigation} or partially~\cite{link2025experimental} replacing hydrocarbon fuels with hydrogen in combustion systems. However, stoichiometric hydrogen flames exhibit higher adiabatic flame temperatures than traditional fuels, leading to increased emissions of harmful nitrogen oxides (NO$_{\rm x}$). To meet emission regulations, recent research has concentrated on studying hydrogen combustion under lean premixed conditions, where lower adiabatic flame temperatures reduce NO$_{\rm x}$ formation across the thermal route~\cite{porcarelli2023suppression,acquaviva2025influence}. Due to its high reactivity and lower heating value, hydrogen is ideal for attaining ultra-lean combustion regimes while avoiding the risk of lean blow-off~\cite{cho2009improvement}. Nevertheless, lean hydrogen flames present challenges such as very high flame speeds~\cite{speth2009using}, auto-ignition phenomena~\cite{hu2019large}, and the onset of thermodiffusive instabilities~\cite{lapenna2023hydrogen}, complicating flame control and flashback prevention.

Within this framework, it is essential to rely on cost-efficient, yet accurate CFD simulations to aid the design of novel and safe lean premixed hydrogen combustion systems. Flamelet-based tabulated-chemistry models have been extensively explored and applied to practical cases. They are based on the assumption that universal local laminar-like structures appear across a turbulent flame front. Due to hydrogen's unique high-diffusivity characteristics, however, existing tabulated-chemistry large eddy simulation (LES) models assuming unity Lewis number (Le) need to be modified to account for differential and preferential diffusion, the possible onset of thermodiffusive instabilities, and for their potential interaction with turbulence~\cite{pitsch2024transition}. 


Various tabulated-chemistry formulations based on two controlling variables, namely progress variable and mixture fraction, have been proposed in the literature to account for differential and preferential diffusion at the resolved scales in laminar conditions. Regele \textit{et al}.~\cite{regele2013two} first introduced a modified mixture fraction equation by including a progress variable dependent source term and considering Le$\neq1$ for the fuel only. This approach was later extended by Schlup and Blanquart~\cite{schlup2019reproducing} to incorporate a full mixture-averaged transport formulation and including thermal diffusion effects.
Another proposed formulation consists of introducing additional terms in the diffusive fluxes of the transported controlling variables 
\cite{de2010inclusion}. This approach was tested only recently with good success to pure hydrogen flames in partially premixed~\cite{perez2025assessment} and fully premixed conditions~\cite{fortes2025analysis}.
Mukundakumar \textit{et al}.~\cite{mukundakumar2021new} proposed a different formulation obtained by inverting the order of the operations performed to reconstruct the diffusive transport terms of the controlling variables, and assuming constant non-unity Lewis numbers. 
Another approach is the composition space method first introduced for unstretched premixed flamelet manifolds~\cite{scholtissek2019selfUnstretched} and then extended to strained and curved premixed flamelets~\cite{scholtissek2019selfStretched}. The extended formulation was tested with good success over spherically expanding lean premixed hydrogen flames at atmospheric~\cite{wen2022flameI} and elevated~\cite{wen2022flameII} pressure, but required a four-dimensional manifold, which is computationally demanding. Reduced manifold formulations have been also attempted for the same 
modelling framework by using unstretched flamelets with varying reactants temperature~\cite{bottler2022flamelet} and curved (unstrained) flamelets~\cite{bottler2023flamelet}.

Further challenges for tabulated-chemistry models of lean hydrogen flames arise at subfilter scales due to the presence of thermodiffusive instabilities. Indeed, differential diffusion effects in sufficiently lean hydrogen/air mixtures lead to intrinsically unstable flames due to thermodiffusive mechanisms~\cite{regele2013two}, as the mixture's effective Lewis number drops below the critical threshold Le$_0$ established analytically by Bechtold and Matalon~\cite{bechtold2001dependence}. Recent studies have addressed these challenges by tabulating data from filtered direct numerical simulations (DNS) performed over the same flame setup~\cite{lapenna2021data}, and introducing an algebraic laminar wrinkling factor to correct the filtered consumption speed to model the effect of subgrid thermodiffusive instabilities over resolved scales~\cite{lapenna2021subgrid}. While this approach was successful in the framework of one-step chemistry simulations~\cite{lapenna2024posteriori}, it still needs improvement in a detailed-chemistry framework~\cite{remiddi2024data}.

The accurate modelling challenges posed by subgrid thermodiffusive instabilities in a large eddy simulation framework become even more pronounced in turbulent flames. Indeed, even if the typical cellular structures are not directly observable due to the instantaneous flame perturbation by turbulent eddies, intrinsic flame instabilities continue to influence flame dynamics. 
Direct numerical simulations~\cite{berger2022synergistic} and experiments~\cite{lapenna2024synergistic} suggested that thermodiffusive instabilities feature a synergistic interaction with turbulence in enhancing hydrogen reactivity and consumption speed. Although these synergistic effects have been proved to progressively vanish with higher and more practical Karlovitz numbers~\cite{berger2024effects}, the flame reactivity still increases due to hydrogen's response to strain triggering a local flame enrichment~\cite{porcarelli2024mitigation}. 

In the context of LES with presumed filtered density functions (FDF), 
extensions 
of the formulations of Regele \textit{et al}.~\cite{regele2013two} and Mukundakumar \textit{et al}.~\cite{mukundakumar2021new} have been proposed with discrete success by Berger \textit{et al}.~\cite{berger2025combustion}, Kai \textit{et al}.~\cite{kai2023flamelet} and Ferrante \textit{et al}.~\cite{ferrante2024differential}, respectively. In these works the reaction rates were obtained from unstretched premixed hydrogen flamelets with varying equivalence ratio, tabulated as a function of progress variable and mixture fraction. However, a recent \textit{a priori} study performed by Böttler \textit{et al}.~\cite{bottler2024can} indicated that a simple two-dimensional unstretched flamelet manifold exhibits limitations in capturing the local thermochemical states in thermodiffusively unstable and turbulent conditions even at the DNS scale. The authors showed that the local error of the tabulated quantities can be significantly reduced through a novel higher-dimensional flamelet manifold consisting of five control variables and considering around 26000 strained and curved flamelets. However, this solution appears very memory-demanding. 

In the present work, a comprehensive \textit{a priori} analysis is performed over an intrinsically unstable laminar and turbulent lean premixed hydrogen flame with differential and preferential diffusion testing a variety of up to two-dimensional manifolds at both unfiltered and filtered grids. The filtered reaction rate is found from the manifolds with the presumed $\beta$-FDF approach~\cite{langella2016large}, with an additional comparison to the one obtained with the F-TACLES approach~\cite{fiorina2010filtered} in unstretched flamelets manifolds. Although further limitations are highlighted for unstretched flamelets manifolds at the filtered grids, two solutions are proposed to significantly reduce the modelling error over a range of filters without increasing the manifold dimension. The first consists of correcting the consumption speed predicted by the unstretched flamelet manifold in the turbulent case with a correction function extrapolated from the laminar \textit{a priori} analysis as a function of the filter width. The second evaluates two strained flamelet approaches: (1) a single counterflow premixed flamelet at a chosen strain rate, and (2) a novel premixed counterflow flamelet manifold at fixed-strain and varying equivalence ratio. Indeed, mean strain rate has been recognised as the main driver of the increased overall flame reactivity experienced in turbulent lean premixed hydrogen flames as it triggers an enrichment of the conditional mean of mixture fraction over the progress variable~\cite{berger2022synergistic}. The capability of strained flamelets to reproduce this shifted profile of mean mixture fraction towards richer states has been already suggested by Berger \textit{et al}.~\cite{berger2024effects} through DNS studies at a bulk Reynolds number of 11000. Unlike the work of Berger \textit{et al}.~\cite{berger2024effects}, however, where the total mean strain rate was solely a function of shear-driven turbulence, this study employs a counterflow premixed flame configuration with a varying applied strain rate~\cite{porcarelli2025stability,fathi2025strain}. This configuration offers two main advantages. First, it allows the extrapolation of the effect of increasing mean strain rate on modelling error and its filter dependence from the laminar flame \textit{a priori} analysis. Second, in the turbulent setting, it enables the isolation of the effects of turbulence-driven mean strain and configuration-driven mean strain, thereby providing insights into their impact on the predictions of the tested models.

This paper is organised as follows. The computational setup and the tested tabulated-chemistry models are discussed in Section~\ref{sec:models}, reporting the setups of the two-dimensional laminar (Section~\ref{sec:models.2D}) and the three-dimensional turbulent (Section~\ref{sec:models.3D}) DNS datasets, the equations of the Gaussian filter applied (Section~\ref{sec:models.filter}), and a complete account of the tabulation models (Section~\ref{sec:models.tabulation}). \textit{A priori} analysis results in terms of local reaction rate error and consumption speed are discussed next in Section~\ref{sec:results}. First, the laminar flame results are reported (Section~\ref{sec:results.lam}). Afterwards, the turbulent flame results are shown including the irreducible error and modelling error (Section~\ref{sec:results.turb}). The choice of the fixed strain rate for the manifold of counterflow flamelets is the discussed in Section~\ref{sec:results.manifoldStrain}. Last, the proposed correction function to improve the prediction of the consumption speed from unstretched flamelet manifolds as a function of the filter width in turbulent flames is discussed (Section~\ref{sec:results.correction}). Finally, conclusions are provided in Section~\ref{sec:conclusions}.

\section{Computational setup and tabulated-chemistry models} \label{sec:models}
A priori analyses are performed using existing laminar two-dimensional data from Porcarelli \textit{et al}.~\cite{porcarelli2025stability} and turbulent three-dimensional data from Fathi \textit{et al}.~\cite{fathi2025strain}. These two simulations are chosen because they both consist of the same strained counterflow configuration with the same inflow conditions (and thus flame setup), and similar ranges of strain rates are established, thus allowing for a meaningful comparison between the laminar and the turbulent cases. The flame setups are described in detail below.

\subsection{2D laminar setup} \label{sec:models.2D}
\begin{figure}
\centering
\includegraphics[width=0.7\textwidth]{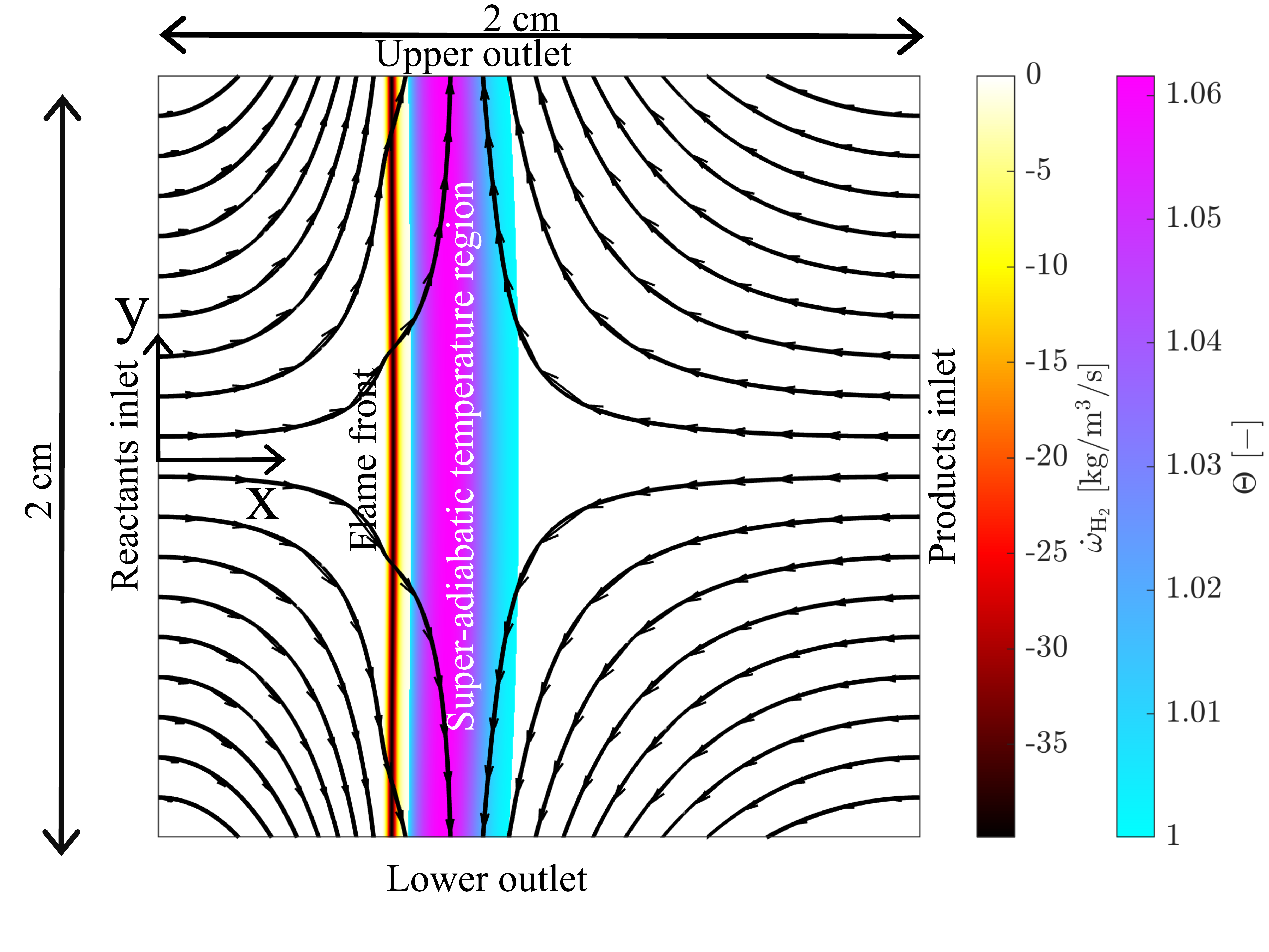}
\caption{Sketch of the two-dimensional laminar reactants-to-products counterflow setup, showing the flame solution at $a=706.85$ s$^{-1}$~\cite{porcarelli2025stability}. The flame front is identified by the contour of hydrogen source term $\dot{\omega}_{\rm H_2}$, and super-adiabatic temperatures are visible in the region of super-unity temperature progress variable $\Theta$.} \label{fig:CF2D}
\end{figure}

The laminar DNS setup consists of a two-dimensional counterflow flame in reactants-to-products configuration as sketched in Figure~\ref{fig:CF2D}. 
Lean conditions are established at an equivalence ratio $\phi= 0.5$ and the reactants temperature and pressure are fixed respectively to $T_r=300$ K and $p=1$ atm. Detailed kinetic data of reactions are taken from the Conaire chemical mechanism~\cite{conaire2004comprehensive}. The nominal applied strain rate is defined as 
\begin{equation}
    \label{eq:a}
    a = \frac{\left | u_r \right | + \left | u_p \right |}{L},
\end{equation}
where $L=2$ cm is the domain length, $u_r$ and $u_p$ are the velocity at the reactants and products boundary, respectively~\cite{langella2016large}. Three levels of nominal strain rates are considered in this study: $a=706.85$ s$^{-1}$, $a=1447.5$ s$^{-1}$, and $a=3633.5$ s$^{-1}$. Further information on the solver and the setup can be found in~\cite{porcarelli2025stability}. At these levels of strain rate, it was shown that thermodiffusive instabilities are suppressed, resulting in a planar flame front once steady state is achieved~\cite{porcarelli2025stability}. The \textit{a priori} analysis is performed for each case at different applied strain rates over a sample steady-state, planar simulation snapshot. Note, however, that variations with respect to the states of an unstretched laminar flame are still present due to the presence of different strain rates, leading to super-adiabatic temperatures (see Fig.~\ref{fig:CF2D}) and super-equilibrium products, thus still challenging the tested models.

\subsection{3D turbulent setup} \label{sec:models.3D}
\begin{figure}
\centering
\includegraphics[width=0.8\textwidth]{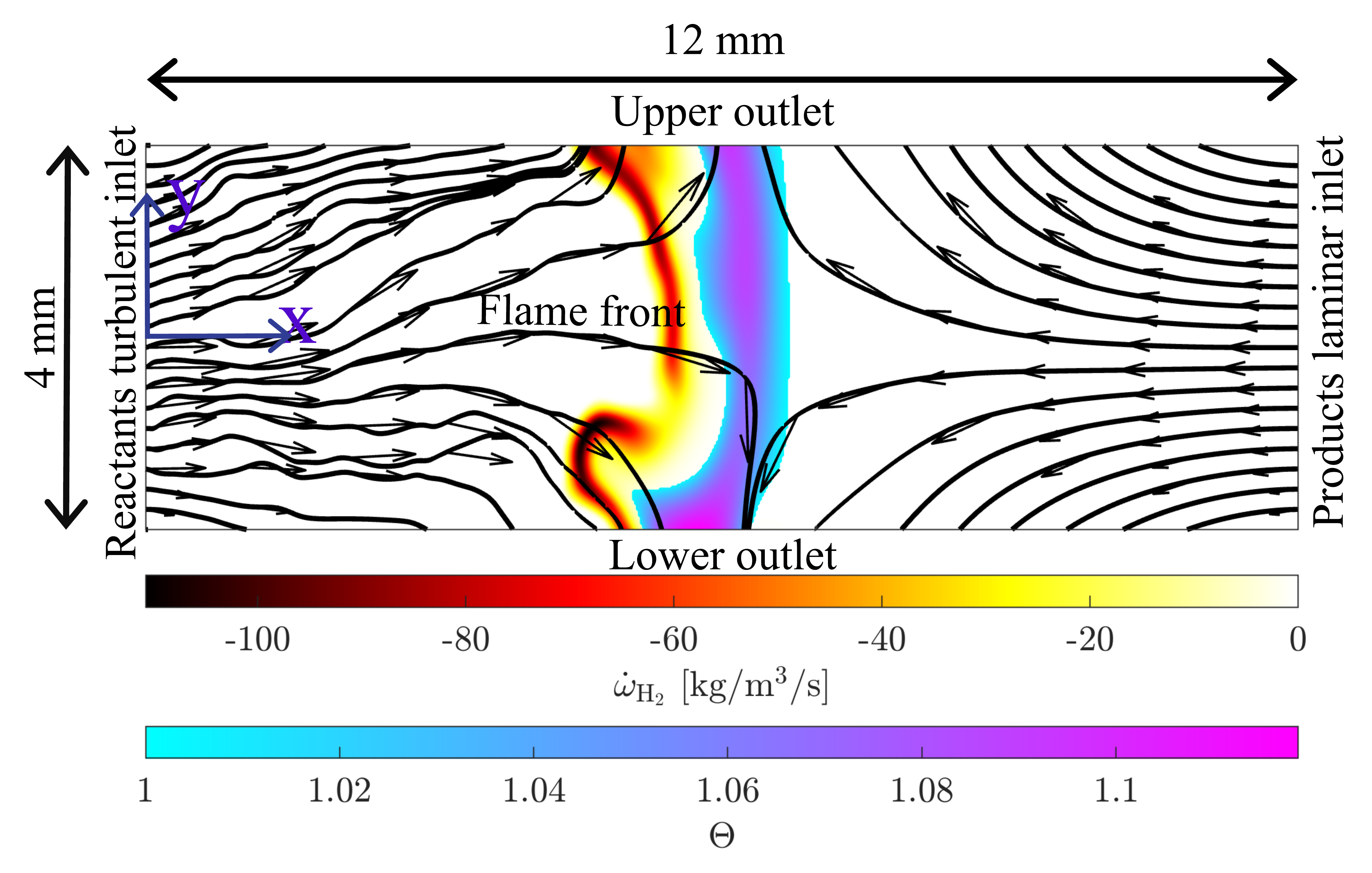}
\caption{Cross-section sketch at the mid cutting plane of the three-dimensional turbulent counterflow setup at a sample timestep, showing a snapshot of the flame solution at $a=2000$ s$^{-1}$~\cite{fathi2025strain}. The flame front is identified by the contour of hydrogen source term $\dot{\omega}_{\rm H_2}$, and super-adiabatic temperatures are visible in the region of super-unity temperature progress variable $\Theta$.} \label{fig:CF3D}
\end{figure}

The turbulent DNS setup consists of a three-dimensional counterflow flame in reactants-to-products configuration. A cross-section of the setup can be observed in Figure~\ref{fig:CF3D}. The setup features periodic boundary conditions over the z axis, so that the complete computational domain should be considered an `extrusion' of the counterflow section reported by $L_z=3$ mm. 
The same equivalence ratio, reactants temperature, and pressure conditions are considered as in the laminar case, while detailed kinetic data of reactions are taken from a combination of the hydrogen skeletal mechanism from Sanchez and Williams~\cite{sanchez2014recent} and the NO$_{\rm x}$ module from Capurso \textit{et al}.~\cite{capurso2023nox}. An inflow boundary with homogeneous isotropic turbulence is superimposed using the digital filter method at the reactants side of the counterflow. The average inflow velocity is kept fixed at all cases, such that a Karlovitz number of ${\rm Ka} = (l_t/\delta_f)^{-1/2}(u^{\prime}/s_L)^{3/2} = 3.143$ is achieved, where $l_t$ is the integral length scale, $\delta_f$ is the flame thickness, $u^{\prime}$ is the root-mean-square velocity, and $s_L$ is the laminar flame speed. The nominal applied strain rate, also referred to as "bulk" strain rate, is changed by adjusting the products inlet velocity (see Eq.~\eqref{eq:a}), whose inflow is kept laminar. This way it is possible to increase the bulk strain rate while keeping unaffected the level of turbulence withstood by the flame. Two different levels of bulk strain rate are considered in this study, a=2000 s$^{-1}$ and a=5000 s$^{-1}$. The \textit{a priori} analysis is performed over an instantaneous field at a sample time step. Given the three-dimensional flame setup counting $\sim$16 millions grid points, a single time step is sufficient to collect statistically representative thermochemical state sampling. For a complete description of the solver and the setup, along with the simulation results analysis, the reader is referred to Fathi \textit{et al}.~\cite{fathi2025strain}. 


\subsection{Filtering of DNS dataset} \label{sec:models.filter}
The DNS dataset is filtered using a spatial convolution with a homogeneous isotropic filter function~\cite{pope2001turbulent}:
\begin{equation}
    \label{eq:filter}
    \overline{\phi} (\bm{x},t) = \int _{-\infty}^{\infty} \phi (\bm{x}-\bm{r},t) G(\bm{r}; \Delta) d\bm{r},
\end{equation}
where the filter function $G$ is taken as a Gaussian filter, and depends on the filter width $\Delta$. 
In order to be consistent with LES formulations, the Favre filtering for a generic quantity $\phi$ is introduced as $\widetilde{\phi} = \overline{\rho \phi} /\overline{\rho}$. 
The Favre-filtering is then applied to progress variable $c$ and mixture fraction $z$, while the density and the reaction rates (in kg/m$^3$/s) are filtered using the conventional Reynolds filter.
A range of filters widths around the laminar flame thickness is applied to the DNS dataset. This range is reported in Table~\ref{tab:filters} for the different cases at different applied strain rates analysed in this work, along with corresponding flame thicknesses. Note that the same ratio $\Delta/\delta_f$ is kept for different cases. Also, the largest filter size used in the laminar cases of $\Delta=4\delta_f$ could not be achieved for the turbulent cases due to the smaller domain size in the latter, leading to a very coarse filtered grid.
\begin{table}
\centering
\caption{Range of filter widths applied to the different DNS datasets with the corresponding laminar flame thicknesses $\delta_f$.}
\begin{tabular}{lccc} \toprule
 \textbf{Case name} & a [s$^{-1}$] & $\Delta$ [mm] & $\delta_f$ [mm] \\ \midrule
 \multicolumn{4}{c}{Laminar flames} \\
 a707L & 706.85 & $\delta_f$,2$\delta_f$,4$\delta_f$ & 0.371 \\
 a1448L & 1447.5 & $\delta_f$,2$\delta_f$,4$\delta_f$ & 0.346 \\
 a3634L & 3633.5 & $\delta_f$,2$\delta_f$,4$\delta_f$ & 0.299 \\ \midrule
 \multicolumn{4}{c}{Turbulent flames} \\
 a2000T & 2000 & $\delta_f$/2,$\delta_f$,2$\delta_f$ & 0.285 \\
 a5000T & 5000 & $\delta_f$/2,$\delta_f$,2$\delta_f$ & 0.239 \\ \bottomrule
 \end{tabular}
\label{tab:filters}
\end{table}

\subsection{Tabulation approach} \label{sec:models.tabulation}
In this work, three different types of flamelet manifolds are considered as follows, and summarised in Table~\ref{tab:flamelets}. 
\begin{itemize}
    \item A single counterflow reactants-to-products strained flamelet at the nominal equivalence ratio $\phi=0.5$ (1DS). On the unfiltered grid, this one-dimensional manifold is parametrised with a progress variable $c$. The applied strain rate in the flamelet is the same as that of the DNS for the 2D laminar cases. In the turbulent cases the strain rate experienced by the flame varies in space and time as it is a combination of applied and turbulence-driven strain. Therefore, the applied strain rate on the flamelet is chosen so that the tangential strain on the flamelet is close to the conditional mean estimated from the DNS~\cite[see][Fig. 7]{fathi2025strain}; 
    %
    \item A manifold of 300 unstretched premixed flamelets with  equivalence ratio varying from $\phi=0.3$ to $\phi=1$ (2DU). On the unfiltered grid, two-dimensional manifold is parametrised with a progress variable $c$ and a mixture fraction $z$. 
    \item A manifold of 300 reactants-to-products counterflow flamelets at the same fixed applied strain rate and with  equivalence ratio varying from $\phi=0.3$ to $\phi=1$ (2DFS). The value of applied strain rate is chosen here as for the 1DS case. 
    On the unfiltered grid, the resulting two-dimensional manifold is parametrised with a progress variable $c$ and a mixture fraction $z$.
\end{itemize}
%
%
\begin{table}
\centering
\caption{Summary of the tested flamelet manifolds.}
{\begin{tabular}{lc} \toprule
 \textbf{Manifold name} & Manifold type \\ \midrule
 1DS & Fixed-strain counterflow flamelet at $\phi=0.5$ \\
 2DU & Unstretched flamelets with varying $\phi$ \\
 2DSF & Fixed-strain counterflow flamelets with varying $\phi$\\ \bottomrule
 \end{tabular}}
\label{tab:flamelets}
\end{table}
Note that while higher dimensional manifolds are possible, they are not considered in the present work where the objective is to ensure 
that potential LES based on these manifolds remain efficient in terms of computational time and memory requirements (note that the manifold dimension will increase in the case of filtered meshes, see Secs.~\ref{sec:models.tabulation.FDF} and~\ref{sec:models.tabulation.FTACLES}). 
\newline

The flamelet manifolds are computed with Chem1D~\cite{chem1d}. In the counterflow cases, the reactants and products velocities are computed from the imposed fixed strain rate $a$ as:
\begin{equation}
    u_r = \frac{aL}{1+\frac{\rho_r}{\rho_p}}, \qquad u_p = u_r \frac{\rho_r}{\rho_p},
\end{equation}
where $L=2$ cm is the domain length and subscripts $r$ and $p$ refer to reactants and products streams respectively.
The progress variable used in the present \textit{a priori} analysis is based on the hydrogen fuel, $c= 1-Y_{\rm H_2}/{Y_{\rm H_2,r}}$, where $Y_{\rm H_2,r}$ denotes the mass fraction of hydrogen in the reactants at the nominal equivalence ratio. This choice guarantees the monotonicity of the progress variable across the flamelet for the lean conditions investigated here, which is not the case for other choices.
The Bilger's definition is used for the mixture fraction~\cite{bilger1990reduced}:
\begin{equation}
    \label{eq:Zbilger}
    z = \frac{\frac{1}{2 W_{\rm H}} (\mathcal{Y}_{\rm H} - \mathcal{Y}_{\rm H,p}) - \frac{1}{W_{\rm O}} (\mathcal{Y}_{\rm O} - \mathcal{Y}_{\rm O,p})}{\frac{1}{2 W_{\rm H}} (\mathcal{Y}_{\rm H,r} - \mathcal{Y}_{\rm H,p}) - \frac{1}{W_{\rm O}} (\mathcal{Y}_{\rm O,r} - \mathcal{Y}_{\rm O,p})},
\end{equation}
where $W_k$ is the molar mass of element $k$, $\mathcal{Y}_{\rm H}$ and $\mathcal{Y}_{\rm O}$ are the elemental mass fractions of monoatomic hydrogen and oxygen, respectively, and the subscripts $r$ and $p$ indicate reactants and products side of the flamelet. 

Additional controlling variables are needed when the DNS dataset is filtered, emulating the subgrid turbulence-flame interaction in a LES. This work mainly evaluates the presumed filtered density function (FDF) approach with $\beta$-FDF, with additional comparisons to the performance of the filtered tabulated-chemistry approach for LES (F-TACLES) in the context of 2DU-type manifolds.
These approaches are discussed next.

\subsubsection{Presumed FDF approach} \label{sec:models.tabulation.FDF}
The performance of different manifolds based on the presumed FDF approach for LES is assessed \textit{a priori}.
The filtered reaction rate in the presumed FDF approach can be written as~\cite{langella2016unstrained}:
\begin{equation}
    \label{eq:omegaTildeComplete}
    \overline{\dot{\omega}} = \overline{\rho}\int _0 ^1\int _0 ^1 \frac{\dot{\omega} (\zeta, \eta)}{\rho (\zeta, \eta)} \widetilde{P}(\zeta, \eta; \widetilde{c}, \widetilde{z}, \sigma^2_{c,\rm sgs}, \sigma^2_{z,\rm sgs}) d\zeta d\eta,
\end{equation}
where $\zeta$ and $\eta$ are the sample space variables of progress variable and mixture fraction, respectively, $\sigma_{c,\rm sgs}=\widetilde{c^2}-\widetilde{c}^2$ and $\sigma_{z,\rm sgs}=\widetilde{z^2}-\widetilde{z}^2$ represent the subgrid variance of $c$ and $z$, respectively, and $\widetilde{P}(\zeta, \eta; \widetilde{c}, \widetilde{z}, \sigma^2_{c,\rm sgs}, \sigma^2_{z,\rm sgs})$ is their joint FDF (sometimes referred to as subgrid probability density function, PDF). In LES of hydrocarbon fuels  
$z$ only describes the local mixing of fuel and oxidizer, and 
$c$ and $z$ can be treated as statistically independent~\cite{Chen03092018}.
However, this is not the case in hydrogen flames, where local fluctuations of mixture fraction across the flame front are present due to 
preferential diffusion. Following the methodology proposed by Berger \textit{et al}.~\cite{berger2025combustion}, it is therefore convenient to introduce a flamelet index $\phi_{\rm FL}$. For every flame state described by $(c,z)$, $\phi_{\rm FL}$ represents the nominal equivalence ratio of the corresponding flamelet in the chosen manifold, so that $\phi_{\rm FL} = f(c,z)$ is a bijective function and can be tabulated from the manifold itself. Unlike the mixture fraction, this quantity is conveniently independent of the progress variable, as it is basically a `label' of each 1D flamelet forming the manifold.
This allows to assume statistical independence between $c$ and $\phi_{\rm FL}$:
\begin{equation}
    \label{eq:statInd}
    \widetilde{P}(\zeta,\xi;\widetilde{c},\widetilde{\phi}_{\rm FL},\sigma^2_{c,\rm sgs},\sigma^2_{\phi_{\rm FL},\rm sgs}) = \widetilde{P}(\zeta;\widetilde{c},\sigma^2_{c,\rm sgs}) \widetilde{P}(\xi;\widetilde{\phi}_{\rm FL},\sigma^2_{\phi_{\rm FL},\rm sgs}),
\end{equation}
where $\xi$ is the sample space variable of the flamelet index. The validity of this assumption has been also confirmed with good approximation from the DNS data of this study (not reported) and previous works (e.g. see~\cite{berger2025combustion}). In the present study, the $\beta$ shape is tested as presumed FDF for the progress variable space, as it showed good performance in the thin reaction zone regime of the Borghi's diagram~\cite{langella2016unstrained} (which is the regime for the turbulent flames here), while other choices are remanded to future studies. 
A delta-function is chosen instead for $\widetilde{P}(\xi)$ since the subgrid variance of mixture fraction is negligible for the cases under investigation. 
Therefore, the tabulated filtered reaction rate is expressed as $\overline{\dot{\omega}}(\widetilde{c},\sigma^2_{c, \rm sgs},\widetilde{\phi}_{\rm FL})$. 

Note that, since $\widetilde{\phi}_{\rm FL}$ is implicitly determined by $\widetilde{z}$, 
\begin{equation}
    \label{eq:zTilde}
    \widetilde{z} = \int _0 ^1 z(\zeta, \widetilde{\phi}_{\rm FL}) \beta(\zeta; \widetilde{c}, \sigma^2_{c,\rm sgs}) d\zeta.
\end{equation}
then the filtered reaction rate in Eq.~\eqref{eq:omegaTildeComplete} still depends on the set of initial parametrisation variables $\widetilde{c}$, $\sigma^2_{c, \rm sgs}$ and $\widetilde{z}$. 
Also, since $\widetilde{z}$ increases monotonically with $\widetilde{\phi}_{\rm FL}$, the latter can be re-interpolated in the functional form $\overline{\dot{\omega}}(\widetilde{c},\sigma^2_{c, \rm sgs},\widetilde{z})$. The importance of this final step, which has been controversially omitted in some of the existing studies~\cite[see][for example]{ferrante2024differential,kai2023flamelet,fortes2025large}, is discussed in Section 3 of the supplementary material.

\subsubsection{F-TACLES approach} \label{sec:models.tabulation.FTACLES}
In this approach the filtered reaction rate in a LES is obtained by spatially pre-filtering the reaction rate from the laminar solution~\cite{fiorina2010filtered,lapenna2021data}, i.e.
\begin{equation}
    \label{eq:omegaTildeFT}
    \overline{\dot{\omega}} = \Xi (\Delta)\overline{\dot{\omega}}^* (\widetilde{c}, \widetilde{z}, \Delta),
\end{equation}
where $\overline{\dot{\omega}}^*$ denotes the pre-filtered laminar reaction rate and $\Delta$ is the chosen LES filter width. To account for the subgrid wrinkling at the filtered scales in the turbulent flames, the subgrid scale wrinkling factor $\Xi$ is introduced, which is found from the filtered DNS data as~\cite{poinsot2005theoretical}:
\begin{equation}
    \label{eq:wrinking}
    \Xi (\Delta) = \frac{\overline{\rho s_d |\nabla c|}}{\rho_u s_L |\nabla \widetilde{c}|},
\end{equation}
where $\rho$ and $s_d=\frac{1}{|\nabla c|}\frac{Dc}{Dt}$~\cite{fathi2025strain} are the local density and displacement speed, respectively, $\rho_u$ is the density of the unburnt reactant mixture, and $s_L$ is the laminar flame speed of the unstretched flamelet at $\phi=0.5$.

Attempts to incorporate counterflow strained flamelets within the F-TACLES approach are documented in literature for non-premixed~\cite{dillon2024controlling} and partially premixed~\cite{dillon2024new} setups. However, they are not applicable in the framework of this work assessing fully premixed counterflow flamelet manifolds. Therefore, this approach is only employed in the context of unstretched flamelet manifolds, which is well-established in literature also for hydrogen flames~\cite{dillon2024new}.
\newline


\section{Results} \label{sec:results}

\subsection{Laminar flames} \label{sec:results.lam}
The 2D DNS dataset of laminar flames is assessed first in order to evaluate the robustness of the tabulation approaches proposed in Section~\ref{sec:models} without turbulence. Note that as the filter size increases, $\sigma^2_{c,sgs}>0$ even in the case the flame is laminar. The full list of cases assessed is provided in Table~\ref{tab:cases}. Additional cases testing the 2DSF-type manifolds (see Table~\ref{tab:flamelets}) can be found in Section 1 of the supplementary material.
\begin{table}
\centering
\caption{Summary of tested models over the laminar setup with nomenclature. Recall that the flamelet strain rate in the 1DS manifolds is the same as the nominal strain rate of the corresponding case.}
{\begin{tabular}{lccccc} \toprule
\textbf{Tested model name} & \begin{tabular}{@{}c@{}}Case nominal\\strain rate [s$^{-1}$]\end{tabular} &  \begin{tabular}{@{}c@{}}Manifold\\type\end{tabular} 
& \begin{tabular}{@{}c@{}}Subgrid\\model \end{tabular} & \begin{tabular}{@{}c@{}}parametrisation\\variables\end{tabular} \\ \midrule
 a707L-1DS-$\beta$FDF & 706.85 &  1DS & $\beta$-FDF & $\widetilde{c}$,$\sigma^2_{c, \rm sgs}$ \\
 a1448L-1DS-$\beta$FDF & 1447.5 &  1DS & $\beta$-FDF & $\widetilde{c}$,$\sigma^2_{c, \rm sgs}$ \\
 a3634L-1DS-$\beta$FDF & 3633.5 &  1DS & $\beta$-FDF & $\widetilde{c}$,$\sigma^2_{c, \rm sgs}$ \\
 
 a707L-2DU-$\beta$FDF & 706.85 &  2DU & $\beta$-FDF & $\widetilde{c}$,$\sigma^2_{c, \rm sgs}$,$\widetilde{z}$ \\
 a1448L-2DU-$\beta$FDF & 1447.5 &  2DU & $\beta$-FDF & $\widetilde{c}$,$\sigma^2_{c, \rm sgs}$,$\widetilde{z}$ \\
 a3634L-2DU-$\beta$FDF & 3633.5 &  2DU & $\beta$-FDF & $\widetilde{c}$,$\sigma^2_{c, \rm sgs}$,$\widetilde{z}$ \\

 a707L-2DU-FT & 706.85 &  2DU & F-TACLES & $\widetilde{c}$,$\widetilde{z}$,$\Delta$ \\
 a1448L-2DU-FT & 1447.5 &  2DU & F-TACLES & $\widetilde{c}$,$\widetilde{z}$,$\Delta$ \\
 a3634L-2DU-FT & 3633.5 &  2DU & F-TACLES & $\widetilde{c}$,$\widetilde{z}$,$\Delta$ \\


 \bottomrule
 \end{tabular}}
\label{tab:cases}
\end{table}

\subsubsection{Fixed-strain flamelet} \label{sec:results.lam.1DS}
\begin{figure}
\centering
\subfloat[1DS manifolds (first three cases in Table~\ref{tab:cases}). \label{fig:L1DS}]{%
\resizebox*{0.48\textwidth}{!}{\includegraphics{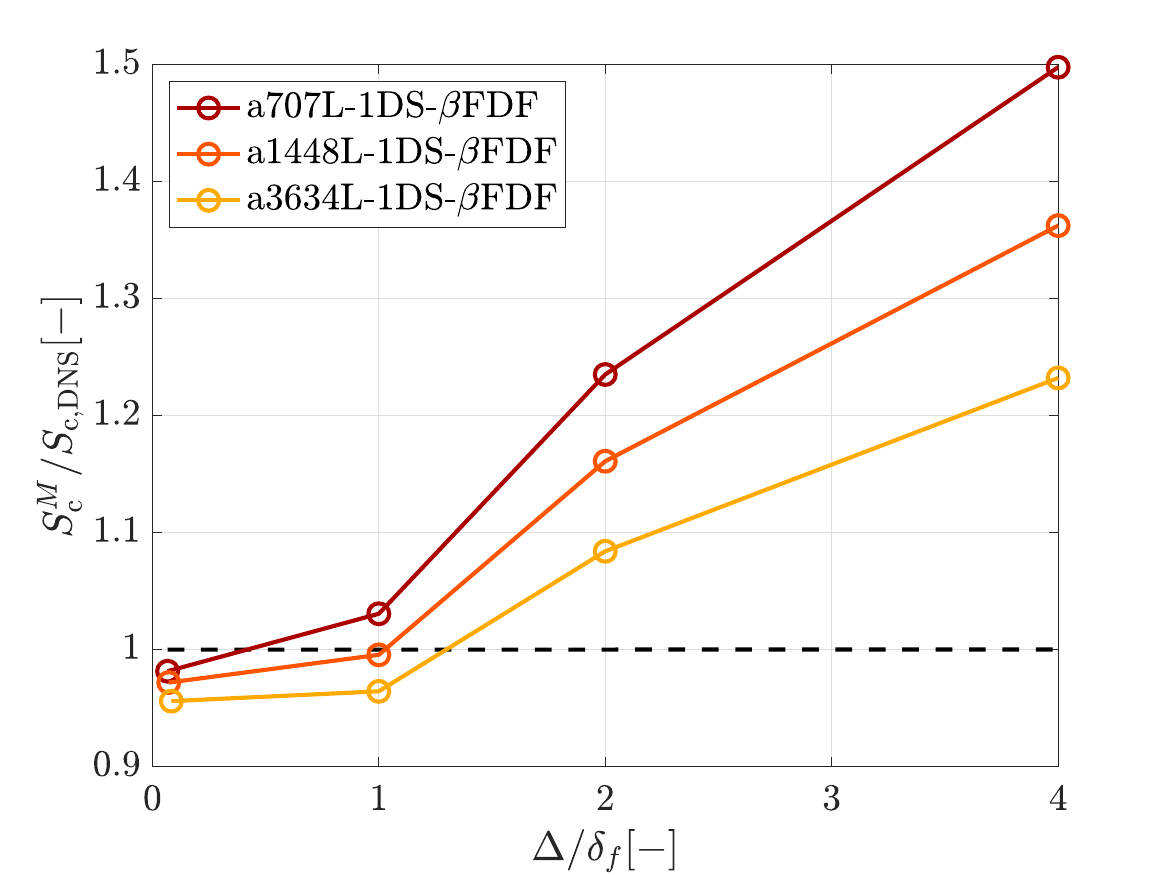}}}\hspace{5pt}
\subfloat[2DU manifolds (cases four to nine in Table\ref{tab:cases}). \label{fig:L2DU}]{%
\resizebox*{0.48\textwidth}{!}{\includegraphics{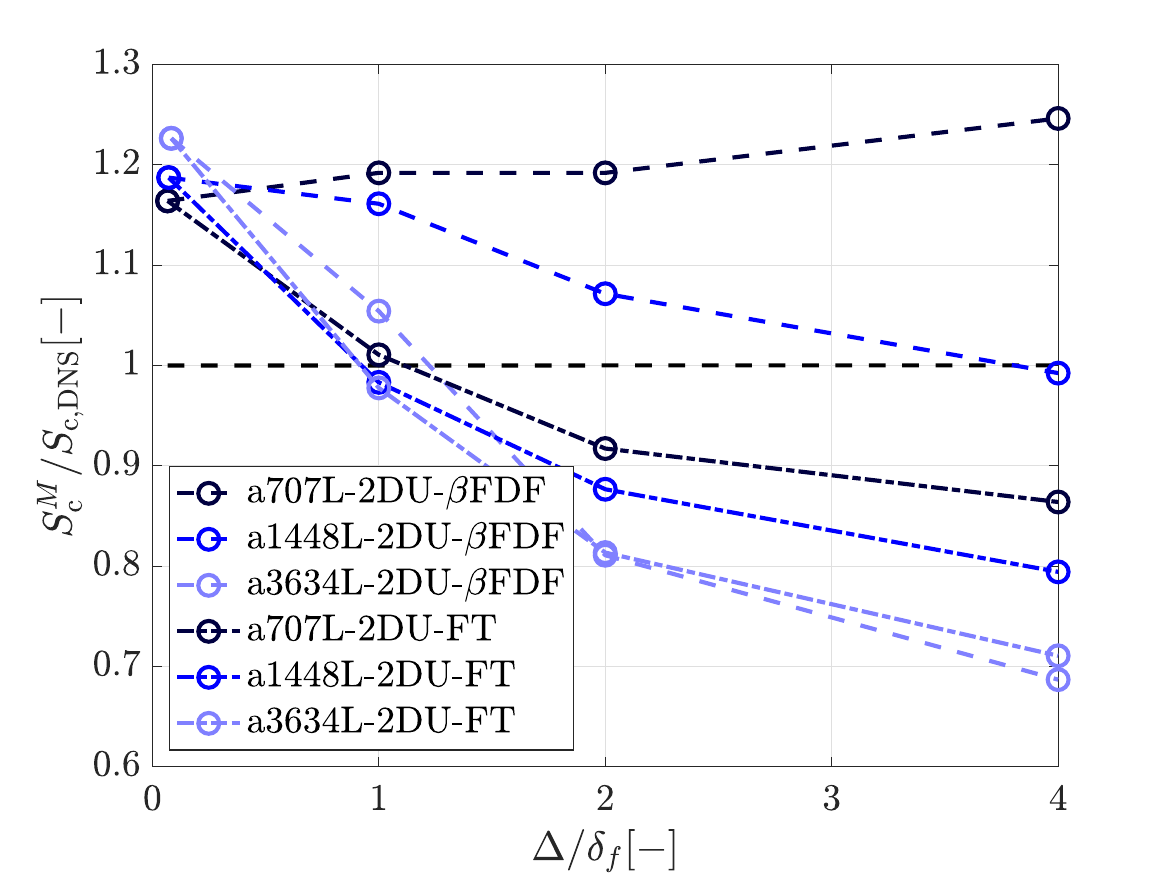}}}
\caption{Ratio of modelled versus DNS consumption speed for the first three cases (a) and cases four to nine (b) of Table~\ref{tab:cases}, for increasing filter widths.} \label{fig:LSc}
\end{figure}

In this section, the ability of the fixed-strain manifold (first three cases in Table~\ref{tab:cases}) to reproduce the states of a two-dimensional laminar and strained lean premixed hydrogen flame is tested. 
The fuel source term $\dot{\omega}_{\rm H_2}^M$, filtered and unfiltered, is reconstructed by entering the flamelet manifold with the controlling variables indicated in Table~\ref{tab:cases}. The modelled consumption speed is thus computed as
\begin{equation}
    \label{eq:ScLam}
    S_c^M = \frac{1}{\rho_u Y_{\rm H_2,u} L_y} \int_A \dot{\omega}_{\rm H_2}^M dA,
\end{equation}
where $L_y=2$ cm is the stream transversal length of the domain, and is compared to the one obtained directly from the DNS data, $S_{c,\rm DNS}$. Results are shown in Figure~\ref{fig:L1DS} for increasing filter widths and three levels of applied strain.
%
%
The graph shows that the relative error is below 5\% for filter sizes up to $\Delta=\delta_f$. This is somewhat expected, since the used flamelet is a one-dimensional representation of the two-dimensional flame at the given strain. The small discrepancies observed are attributed to the numerical errors in the different solvers used. 
The error rapidly increases for filter sizes $\Delta>\delta_f$, which is due to the limitations of the $\beta$-FDF in mimicking the asymptotic bimodal behaviour at large filter sizes~\cite{nilsson2019priori}. Interestingly, this overestimate appears to be reducing at increasing strain rates, the reasons for which are discussed in Section 2 of the Supplementary Material. 

\subsubsection{Unstretched flamelets} \label{sec:results.lam.2DU}
In this section, the ability of a manifold of unstretched flamelets at varying equivalence ratio to reproduce the states of the two-dimensional laminar and strained lean premixed hydrogen flame front is assessed for three levels of applied strain (fourth to ninth case in Table~\ref{tab:cases}). For this manifold type, we compare results from the presumed FDF to those of the F-TACLES subgrid model to distinguish whether the observed errors stem from parametrisation inaccuracies or from the subgrid closure itself.
The results are shown in Figure~\ref{fig:L2DU}. 
%
%
Unlike the cases with fixed-strain manifold discussed in Section~\ref{sec:results.lam.1DS}, the modelled consumption speed shows a 20\% overestimate already on the unfiltered mesh ($\Delta \rightarrow 0$). This indicates that even if the correct variation of $z$ due to preferential diffusion and stretch is predicted in the reacting flow, thermochemical states of strained flames are not well represented by unstretched manifolds even in laminar conditions.
Except for the case at lowest applied strain and presumed FDF approach, the error on filtered meshes overall tends to decrease for increasing filter sizes regardless of whether using the $\beta$ presumed FDF or the F-TACLES approach. Although this might seem as an improvement at moderate filter sizes, for all these cases the ratio between modelled and DNS consumption speed becomes eventually smaller than unity and keeps decreasing as the filter size increases. This suggests that the small error found for $\Delta \approx \delta_f$, in particular for the F-TACLES cases, is rather due to compensation of effects. 
Note also that the ratio of consumption speeds in the cases of the $\beta$-FDF becomes negative for the two highest applied strain cases despite the known tendency of the $\beta$-FDF methods to overestimate $S_c$, suggesting that the errors introduced by the unstretched thermochemical states play a stronger role. 

\begin{figure}
\centering
\subfloat[Unfiltered fields and filtered fields with $\beta$-FDF. \label{fig:L2DUBPDFCont}]{%
\resizebox*{0.555\textwidth}{!}{\includegraphics{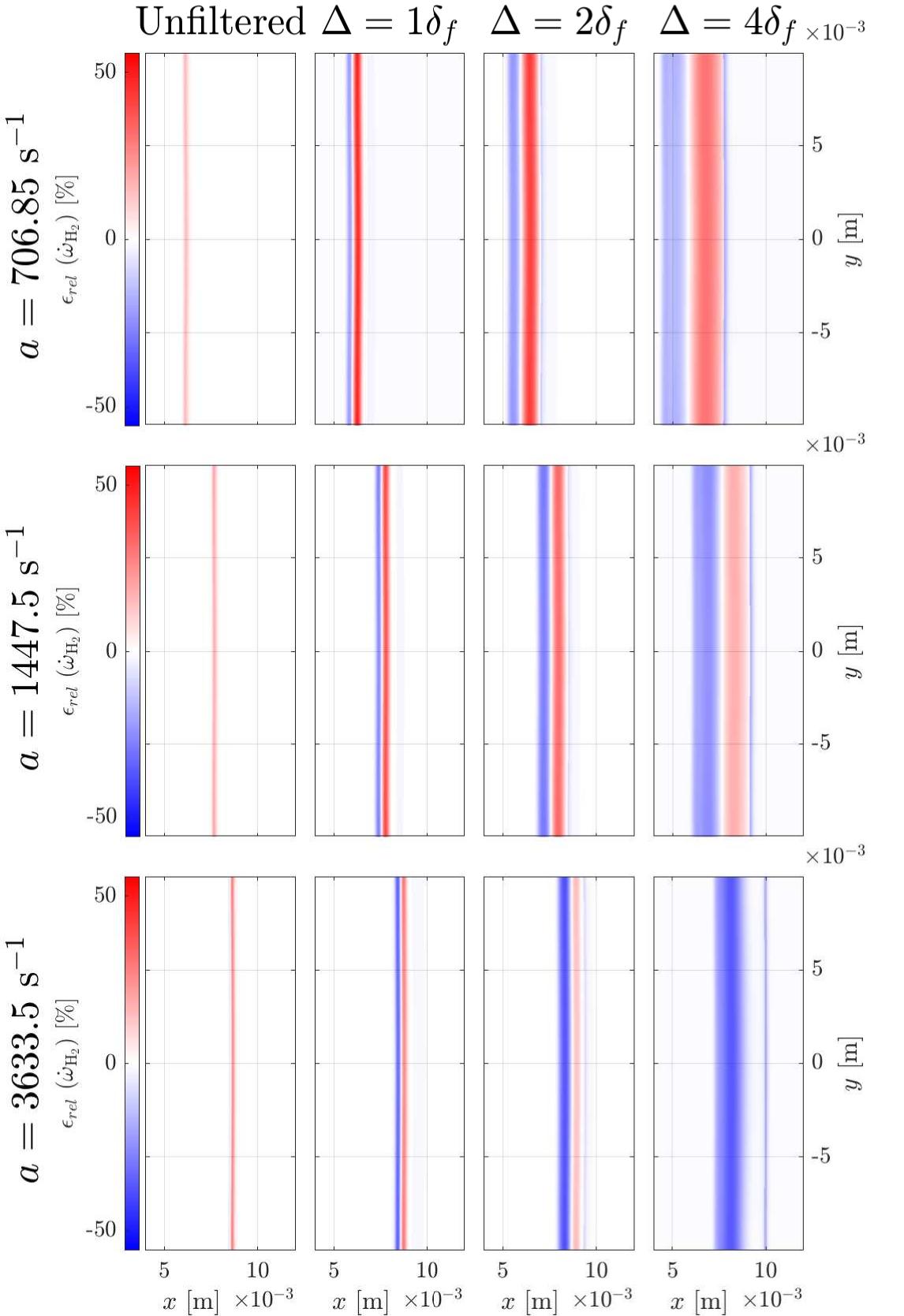}}}\hspace{5pt}
\subfloat[Filtered fields with F-TACLES. \label{fig:L2DUFTCont}]{%
\resizebox*{0.404\textwidth}{!}{\includegraphics{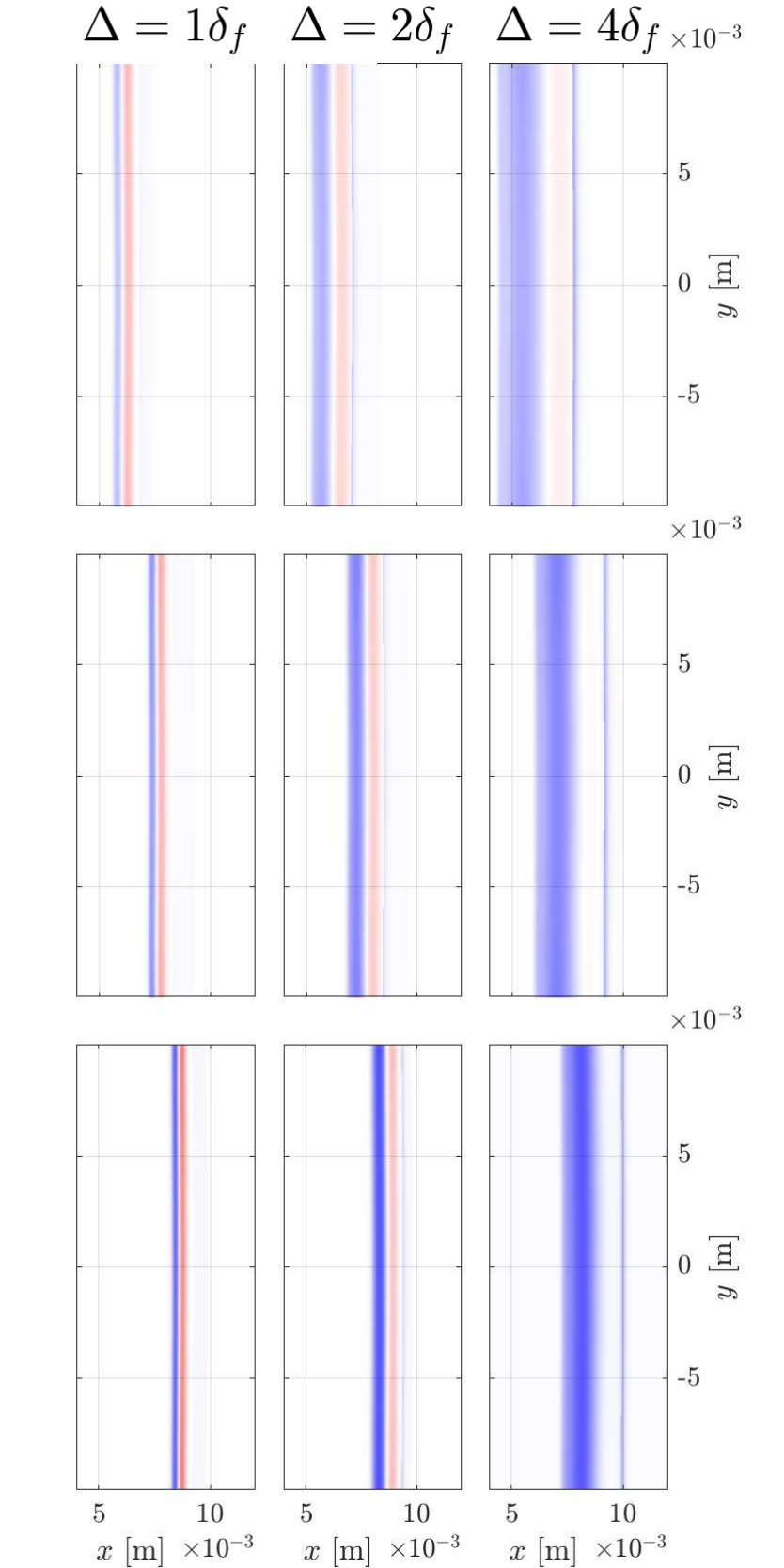}}}
\caption{Relative error of the H$_2$ source term $\dot{\omega}_{\rm H_2}$ over the resolved grid and with increased filter width obtained with the 2DU-type manifolds. The filtered fields are reconstructed with the $\beta$-FDF (a) and with the F-TACLES (b) subgrid models.} \label{fig:L2DUCont}
\end{figure}

In order to understand what regions of space across the flame contribute more to the discrepancies on the consumption speed, the relative error between the fuel reaction rate from the DNS (unfiltered, $\dot{\omega}_{\rm H_2}^{\rm ref}$, or filtered, $\overline{\dot{\omega}}_{\rm H_2}^{\rm ref}$) and that reconstructed from the manifold, is shown in Figure~\ref{fig:L2DUCont}. The relative error is computed as
\begin{equation}
    \label{eq:epsRel}
    \epsilon_{rel} (\dot{\omega}_{\rm H_2}) = \frac{\dot{\omega}_{\rm H_2}^{M} - \dot{\omega}_{\rm H_2}^{\rm ref}}{\rm max \left ( \dot{\omega}_{\rm H_2}^{\rm ref} \right )}.
\end{equation}
For the unfiltered mesh, the overestimate of the reaction rate is stronger in the region of higher progress variable, which corresponds to the region of super-equilibrium products, where the mixture fraction in a lean strained flame also goes above its nominal value~\cite{porcarelli2024mitigation}. This suggests that within an unstretched flamelet manifold, the mixture fraction variations, which correspond to transitions toward richer flamelets, fail to accurately capture the strain-induced reaction rate changes, resulting in an overestimation of reaction rate. Similar considerations can be extended to the reaction rate changes in a positively-curved flame front, which exhibit an overshoot of $z$ at high progress variables similar to that induced by strain.

For filtered meshes ($\Delta>0$) a similar overestimate of the modelled reaction rate is observed for large values of progress variable. 
However, an underestimation is also observable for both presumed FDF and F-TACLES approaches at low values of progress variables, corresponding to regions where the mixture fraction falls below the nominal value in lean strained flames~\cite{porcarelli2024mitigation}. 
These regions of underestimation of reaction rate compensate for the overestimation in the region of super-equilibrium (and superadiabaticity), explaining the relatively small errors on the consumption speed at moderate filter sizes observed in Figure~\ref{fig:L2DU}. 
The region of underestimate becomes dominant at larger filter sizes and higher strain rates, leading to larger and larger underestimation of the consumption speed. Since the same behaviour is observed for both presumed FDF and F-TACLES models, this suggests that the error stems from fundamental parametrisation limitations at filtered grids rather than subgrid model inaccuracies. Moreover, these errors, together with the resulting consumption speed, appear to be filter dependent.



Overall, the \textit{a priori} analysis on the laminar flame showed the limitations of manifolds made of unstretched flamelets in mimicking the local changes of reaction rate due to differential/preferential diffusion, particularly at filtered grids and regardless of the subgrid model used. This suggests that a subfilter correction is necessary in turbulent LES settings to correctly predict the flame properties, which is discussed next. Strained flamelets, on the other hand, showed improved performance in the limit of high strain, giving scope to further tests in turbulent settings.

\subsection{Turbulent flames} \label{sec:results.turb}
The 3D DNS dataset of turbulent flames is now assessed in order to evaluate how the tabulation approaches proposed in Section~\ref{sec:models} perform when the flame is turbulent. The full list of cases assessed is provided in Table~\ref{tab:casesTurb}. Note that the nominal applied strain rate in the DNS differs from the one imposed in the manifold because turbulent eddies induce a mean (positive) tangential strain on the flame~\cite{fathi2025strain}, therefore the total amount of strain experienced by the flame increases. This amount was directly estimated from the DNS. However, since the strain on the flame changes in space and time in the turbulent cases, multiple manifold strain levels have been tested for each DNS case, and further considerations are provided in Section~\ref{sec:results.manifoldStrain}.
\begin{table}
\centering
\caption{Summary of the cases analysed and manifold strategy used for the turbulent strained flames.}
{\begin{tabular}{lccccccc} \toprule
\textbf{Tested model name} & \begin{tabular}{@{}c@{}}Case nominal\\strain rate [s$^{-1}$]\end{tabular} & \begin{tabular}{@{}c@{}}Manifold\\type\end{tabular} & \begin{tabular}{@{}c@{}}Manifold strain \\rate $a$ [s$^{-1}$]\end{tabular} & \begin{tabular}{@{}c@{}}Subgrid\\model \end{tabular} & \begin{tabular}{@{}c@{}}parametrisation\\variables\end{tabular} \\ \midrule

  a2000T-1DSa5000-$\beta$FDF & 2000 &  1DS & 5000 & $\beta$-FDF & $\widetilde{c}$,$\sigma^2_{c, \rm sgs}$ \\
 a2000T-1DSa10000-$\beta$FDF & 2000 &  1DS & 10000 & $\beta$-FDF & $\widetilde{c}$,$\sigma^2_{c, \rm sgs}$ \\

   a5000T-1DSa5000-$\beta$FDF & 5000 &  1DS & 5000 & $\beta$-FDF & $\widetilde{c}$,$\sigma^2_{c, \rm sgs}$ \\
 a5000T-1DSa10000-$\beta$FDF & 5000 &  1DS & 10000 & $\beta$-FDF & $\widetilde{c}$,$\sigma^2_{c, \rm sgs}$ \\

   a2000T-2DU-$\beta$FDF & 2000 &  2DU & - & $\beta$-FDF & $\widetilde{c}$,$\sigma^2_{c, \rm sgs}$,$\widetilde{z}$ \\
 a5000T-2DU-$\beta$FDF & 5000 &  2DU & - & $\beta$-FDF & $\widetilde{c}$,$\sigma^2_{c, \rm sgs}$,$\widetilde{z}$ \\

   a2000T-2DSFa5000-$\beta$FDF & 2000 &  2DSF & 5000 & $\beta$-FDF & $\widetilde{c}$,$\sigma^2_{c, \rm sgs}$,$\widetilde{z}$ \\
 a2000T-2DSFa10000-$\beta$FDF & 2000 &  2DSF & 10000 & $\beta$-FDF & $\widetilde{c}$,$\sigma^2_{c, \rm sgs}$,$\widetilde{z}$ \\
 a2000T-2DSFa15000-$\beta$FDF & 2000 &  2DSF & 15000 & $\beta$-FDF & $\widetilde{c}$,$\sigma^2_{c, \rm sgs}$,$\widetilde{z}$ \\
 
   a5000T-2DSFa5000-$\beta$FDF & 5000 &  2DSF & 5000 & $\beta$-FDF & $\widetilde{c}$,$\sigma^2_{c, \rm sgs}$,$\widetilde{z}$ \\
 a5000T-2DSFa10000-$\beta$FDF & 5000 &  2DSF & 10000 & $\beta$-FDF & $\widetilde{c}$,$\sigma^2_{c, \rm sgs}$,$\widetilde{z}$ \\
 a5000T-2DSFa15000-$\beta$FDF & 5000 &  2DSF & 15000 & $\beta$-FDF & $\widetilde{c}$,$\sigma^2_{c, \rm sgs}$,$\widetilde{z}$ \\

   a2000T-2DU-FT & 2000 &  2DU & - & F-TACLES w/o $\Xi$ & $\widetilde{c}$,$\widetilde{z}$,$\Delta$ \\
 a5000T-2DU-FT & 5000 &  2DU & - & F-TACLES w/o $\Xi$ & $\widetilde{c}$,$\widetilde{z}$,$\Delta$ \\

    a2000T-2DU-FTWF & 2000 &  2DU & - & F-TACLES w $\Xi$ & $\widetilde{c}$,$\widetilde{z}$,$\Delta$ \\
 a5000T-2DU-FTWF & 5000 &  2DU & - & F-TACLES w $\Xi$ & $\widetilde{c}$,$\widetilde{z}$,$\Delta$ \\
 
 \bottomrule
 \end{tabular}}
\label{tab:casesTurb}
\end{table}

\subsubsection{Irreducible error} \label{sec:results.turb.irrErr}
\begin{figure}
    \centering
    \includegraphics[width=0.8\textwidth]{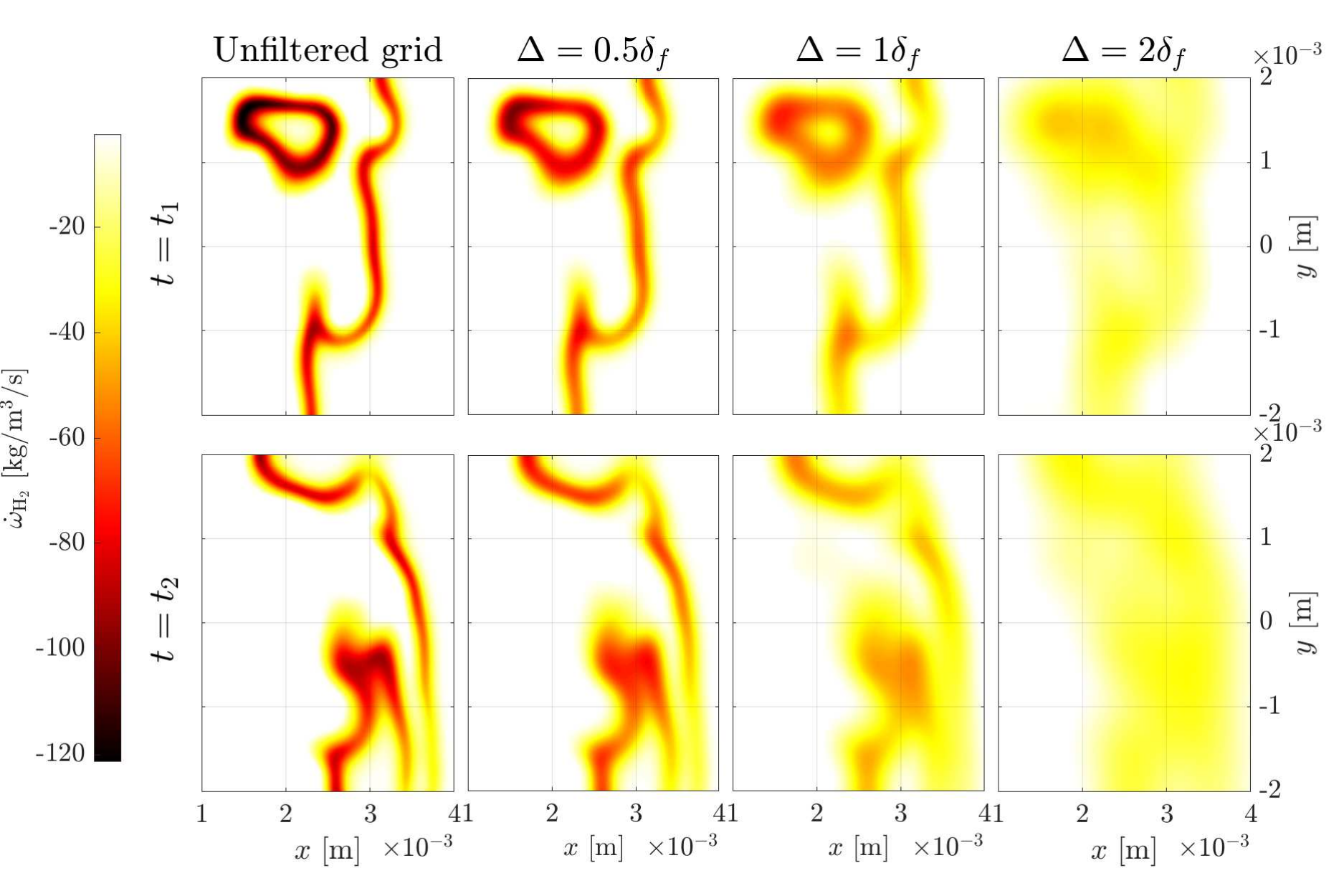}
    \caption{Snapshots at $t=t_1$ (first row) and $t=t_2$ (second row) of H$_2$ source term at the mid-plane of the a5000T simulation over the unfiltered field and with increasing filter width.}
    \label{fig:a5000TcontOmega}
\end{figure}

\begin{figure}
\centering
\subfloat[Unfiltered grid. \label{fig:irrErrorDNS}]{%
\resizebox*{0.23\textwidth}{!}{\includegraphics{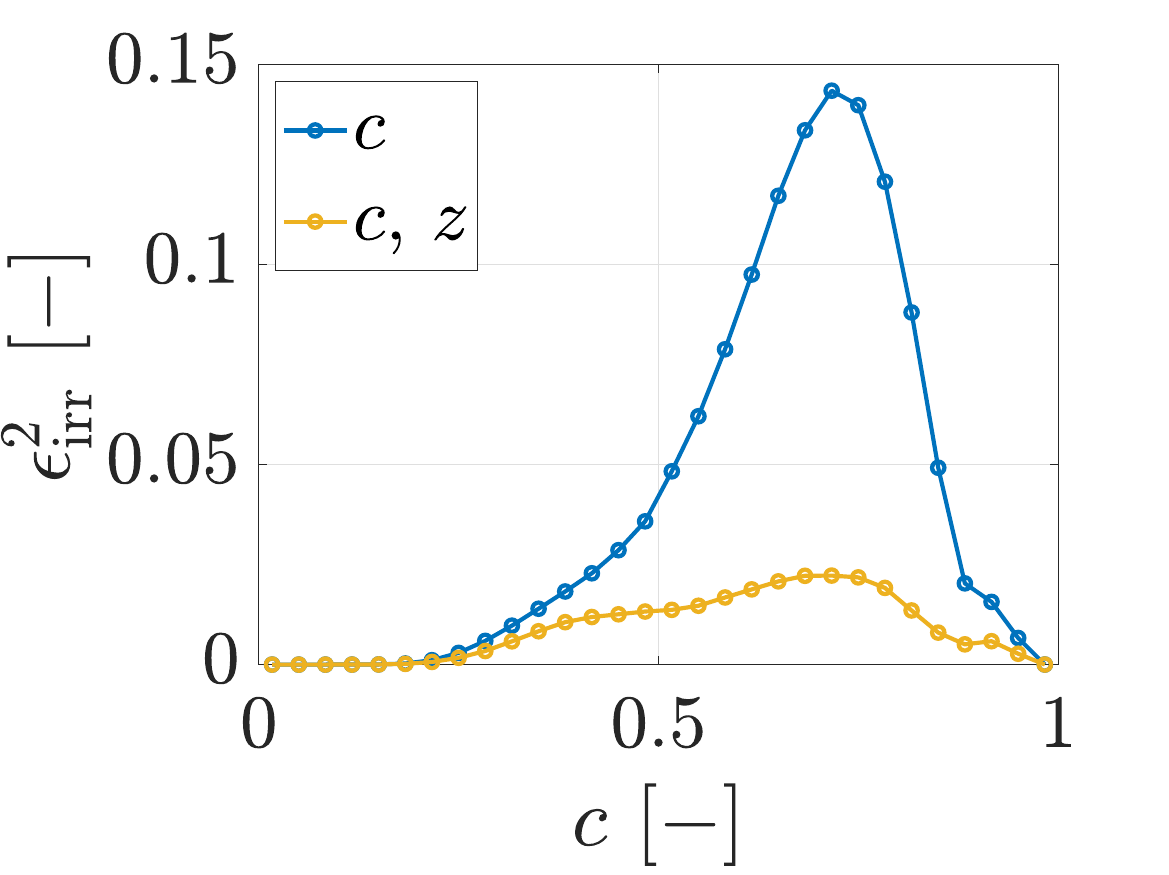}}}\hspace{5pt}
\subfloat[$\Delta=0.5\delta_f$. \label{fig:irrError05}]{%
\resizebox*{0.23\textwidth}{!}{\includegraphics{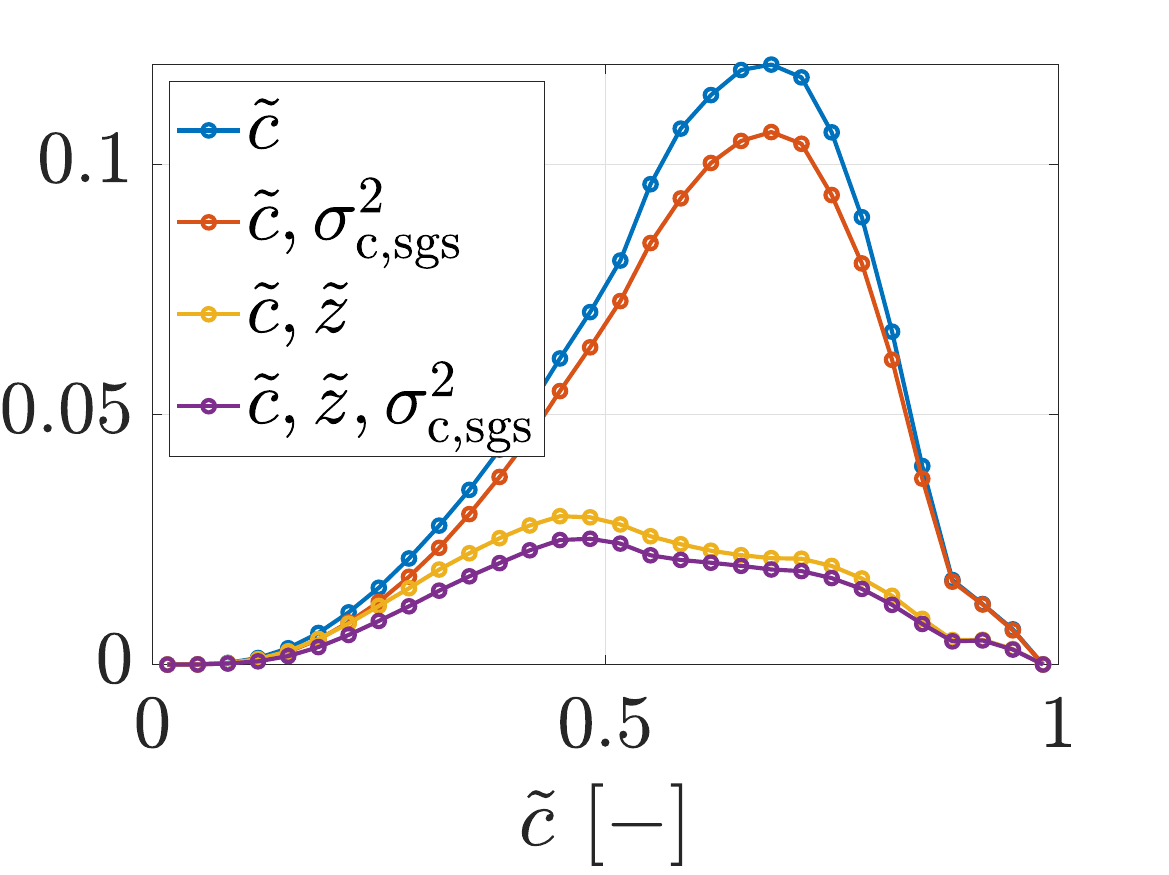}}}\hspace{5pt}
\subfloat[$\Delta=\delta_f$. \label{fig:irrError1}]{%
\resizebox*{0.23\textwidth}{!}{\includegraphics{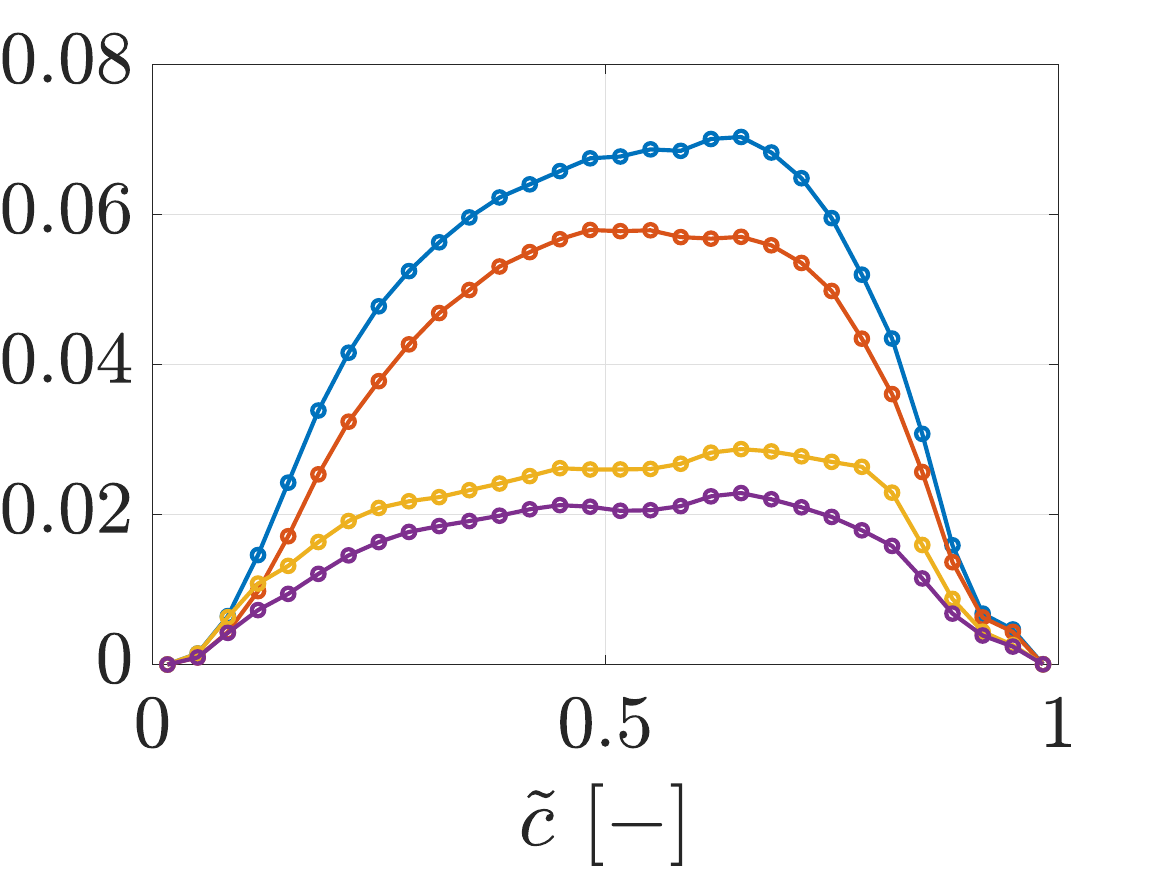}}}\hspace{5pt}
\subfloat[$\Delta=2\delta_f$. \label{fig:irrError2}]{%
\resizebox*{0.23\textwidth}{!}{\includegraphics{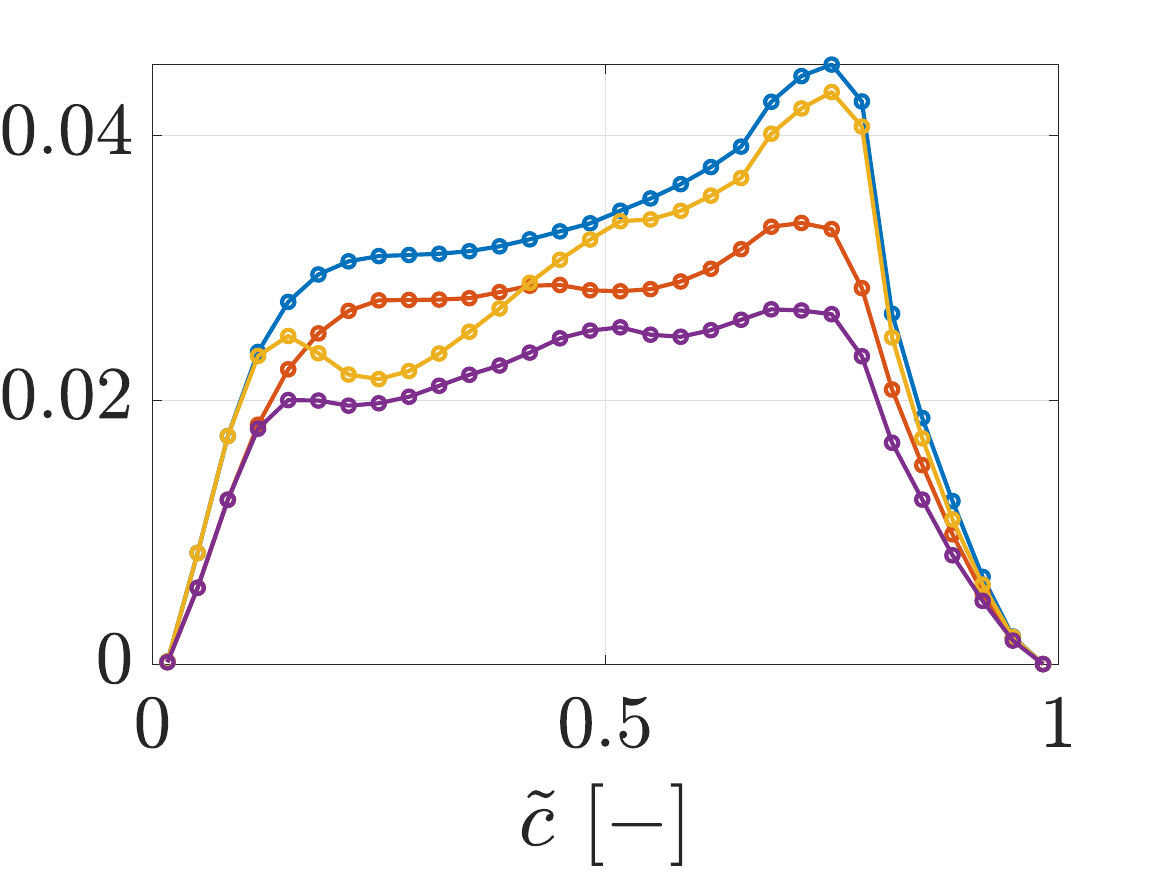}}}
\caption{Irreducible error of the H$_2$ source term $\dot{\omega}_{\rm H_2}$ over the unfiltered grid (a) and with increased filter width (b-d) for the case a5000T. Points are sampled at both $t=t_1$ and $t=t_2$.} \label{fig:irrError}
\end{figure}

An irreducible error is introduced to evaluate how accurately a given set of input parameters can be employed to reproduce
different flame states (identified by the fuel reaction rate in the present analysis).  
The irreducible error is defined as~\cite{moreau2006optimal,berger2018numerically}
\begin{equation}
    \label{eq:irrError}
    \epsilon^2_{irr} = \frac{\left \langle (\dot{\omega}_{\rm H_2} -  \langle \dot{\omega}_{\rm H_2}|\mathbf{\Psi} \rangle ) ^2 | c\right \rangle}{\rm max (\langle \dot{\omega}_{\rm H_2} | c \rangle )^2},
\end{equation}
where $\mathbf{\Psi}$ is a set of input parameters, and quantifies the fluctuations of reaction rate with respect to its mean value. Only for this case, the irreducible error is calculated by sampling DNS data points at two different time steps (nominally $t=t_1$ and $t=t_2$) to improve the accuracy. Figures~\ref{fig:a5000TcontOmega} and~\ref{fig:irrError} show respectively midplane reaction rate contours at the two time steps, and the corresponding irreducible errors, for both unfiltered data and filtered data with increasing $\Delta$. Only the case at nominal applied strain rate $a=5000 \, {\rm s^{-1}}$ is reported as the lower strain rate case yields similar outcomes.

For the unfiltered mesh (Figure~\ref{fig:irrErrorDNS}), using only the progress variable as controlling parameter results in $\dot{\omega}_{\rm H_2}$ fluctuations that reach 15\% of the mean value at $c=0.7$. Although this value is smaller than those reported in existing leaner and low-strain cases (e.g. see $\phi=0.4$ in Figure 2 of~\cite{berger2025combustion}), where the error reaches up to 70\%, it still indicates substantial reaction rate fluctuations due to the combination of differential/preferential diffusion effects and local strain and curvature induced by turbulence. When mixture fraction is added to the controlling parameters the maximum irreducible error on the unfiltered mesh reduces to as low as 2\%. This suggests that, unlike the planar laminar case (see Section 1 in the supplementary material), a manifold made of fixed-strain flamelets with varying equivalence ratio featuring $z$ as additional controlling variable is expected to strongly improve the performance of tabulated chemistry models in the turbulent setting.

When the mesh is filtered, combinations of input parameters including also the progress variable variance are considered. From Figures~\ref{fig:irrError05}-\ref{fig:irrError2}, it can be observed that the error remains always below 5\% when considering a manifold parametrised with $\widetilde{c}$ and $\widetilde{z}$, and it decreases further below 3\% when 
introducing $\sigma^2_{c, \rm sgs}$ among the controlling variables (presumed-FDF approach), although this introduction has a moderate impact at moderate filter sizes.
This suggests that subgrid models not considering the variance as a parameter, like the F-TACLES model, may perform slightly worse solely due to the reduced accuracy of the parametrisation, particularly at coarser grids. 

Interestingly, the error reduction due to including mixture fraction among the controlling variables becomes less strong as the filter width increases. This is due to the fact that mixture fraction variation are solely caused by preferential diffusion effects in the studied flames, and filtering the results reduces mixture fraction gradients thus reducing the reaction rate scatter across the mixture fraction space. 

\subsubsection{Error on consumption speed} \label{sec:results.turb.sc}
\begin{figure}
\centering
\subfloat[Case a2000T. \label{fig:ScTa2000}]{%
\resizebox*{0.48\textwidth}{!}{\includegraphics{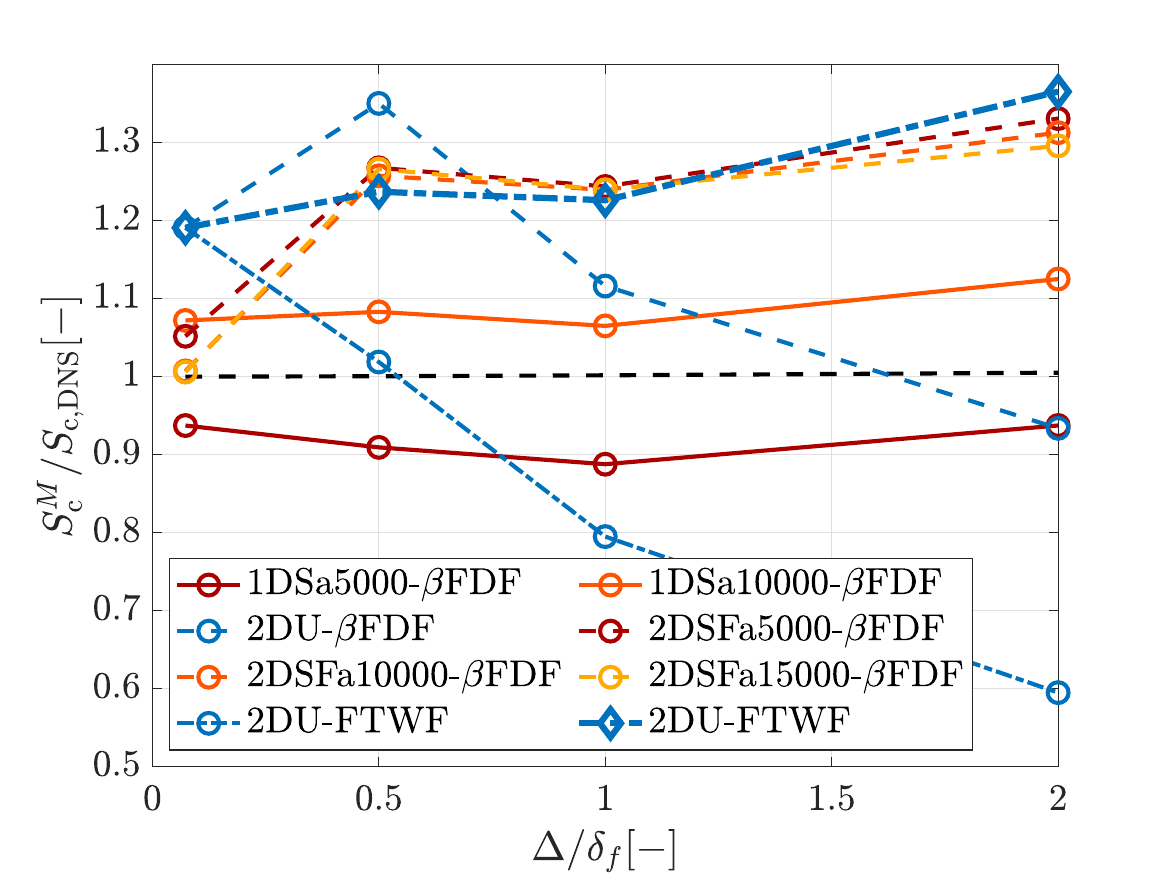}}}\hspace{5pt}
\subfloat[Case a5000T. \label{fig:ScTa5000}]{%
\resizebox*{0.48\textwidth}{!}{\includegraphics{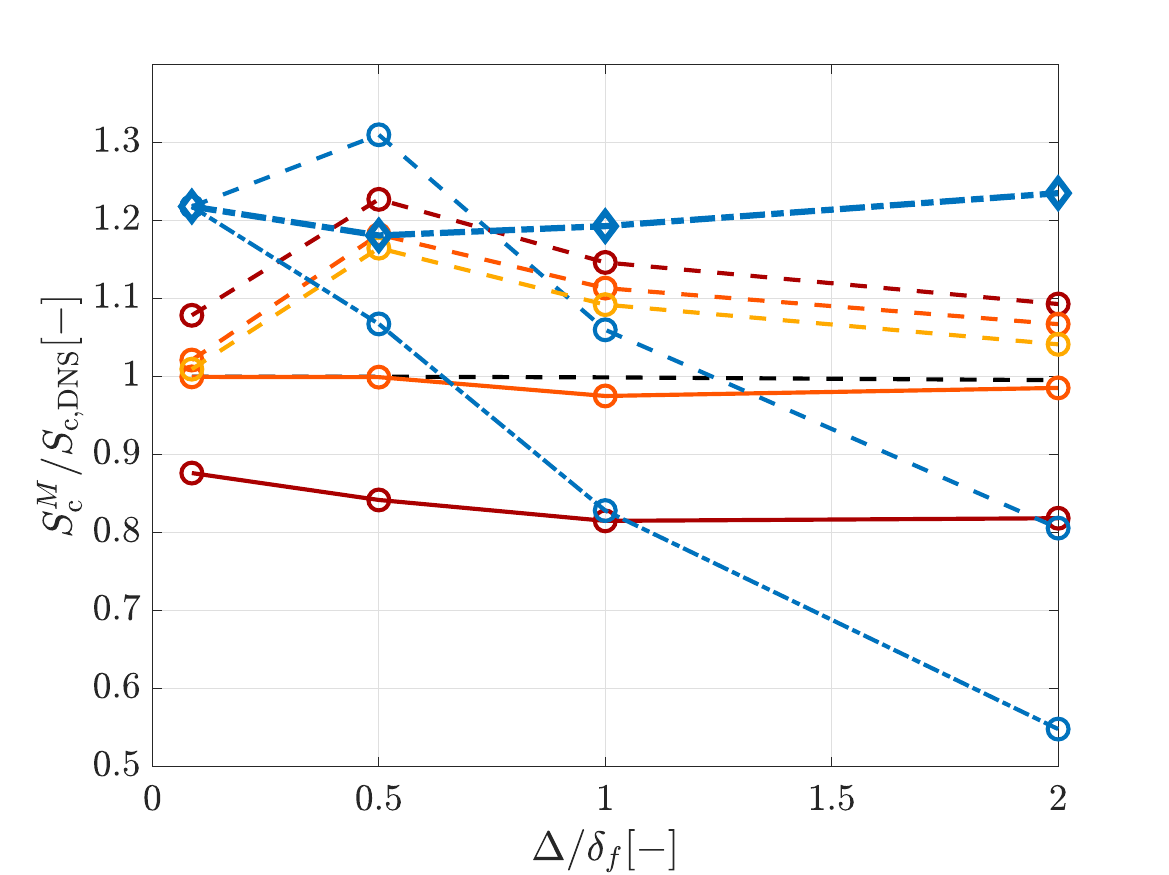}}}
\caption{Ratio of modelled versus DNS consumption speed obtained using the manifold approaches of Table~\ref{tab:casesTurb}, for increasing
filter widths and for values of nominal applied strain of
$a=2000\, {\rm s^{-1}}$ (a) and $a=5000\, {\rm s^{-1}}$ (b).} \label{fig:ScT}
\end{figure}

In this section, the ability of the different manifold approaches of Table~\ref{tab:casesTurb} to predict the consumption speed of a turbulent and strained lean premixed hydrogen flame is assessed. 
The consumption speed is computed as:
\begin{equation}
    \label{eq:ScTurb}
    S_c^M = \frac{1}{\rho_u Y_{\rm H_2,u} L_y L_z} \int_V \dot{\omega}_{\rm H_2} dV,
\end{equation}
where $L_y=4$ mm and $L_z=3$ mm are the vertical and transversal length of the turbulent domain, respectively. The results presented in Figures~\ref{fig:ScT} show data for the cases with nominal applied strain of $a=2000\, {\rm s^{-1}}$ and $a=5000\, {\rm s^{-1}}$.
The manifolds based on unstretched flamelets with varying equivalence ratios (blue lines in Figure~\ref{fig:ScT}) lead to an overestimation of consumption speed of about 20\% on the unfiltered grid, similarly to the laminar cases (see Sec.~\ref{sec:results.lam.2DU}). Furthermore, the strong filter dependence observed in the laminar cases is also observed for the turbulent cases presented here, when the presumed FDF approach with $\beta$-FDF is used. This dependence is instead somewhat relaxed when the F-TACLES approach is used with a wrinkling factor, which is computed in this study directly from the DNS dataset using Eq.~\eqref{eq:wrinking}.
In particular, the modelled consumption speed is overestimated up to 40\% at small filter sizes and underestimated at large filter sizes when the presumed-FDF approach is used. When the F-TACLES approach (without wrinkling factor) is used, on the other hand, the consumption speed is underestimated at all filter sizes by up to 50\% due to the absence of a subgrid wrinkling factor. 
Introducing the wrinkling factor still results in a remaining systematic overestimation of above 20\%. Overall, these results indicate that manifolds based on unstretched flamelets introduce significant errors on consumption speed, which would be difficult to minimise due to the observed dependency on filter width.

Let's now consider the manifolds based on a fixed-strain flamelet (1DS in Table~\ref{tab:casesTurb}). The first observation is that in this case the consumption speed ratio is observed to be mainly independent of the filter width. The quantitative error depends on chosen value for the fixed-strain rate on the flamelet, which as explained earlier varies on the turbulent flame due to the effects of turbulent eddies. For this reason, for each nominal applied strain rate in the turbulent case, two manifolds are tested, respectively with imposed strain on the flamelet of $a = 5000\, {\rm s^{-1}}$ and $a = 10000\, {\rm s^{-1}}$, which are representative of the actual values observed from the DNS data. As observed in Figure~\ref{fig:ScT} the error on the consumption speed is observed to be below or slightly above 10\% for the lower strain case, and almost zero for the higher strain case when flamelets at $a = 10000\, {\rm s^{-1}}$ are used. This is somewhat expected, 
since, the total strain rate experienced by the lower- and higher-strain flames in the DNS are approximately 7000-8000 s$^{-1}$ and 10000-11000 s$^{-1}$, respectively. Nevertheless, results also show that mispredicting the value of strain rate to apply on the flamelet up to 50\% (e.g. see Fig.~\ref{fig:ScTa5000}, moving from $a = 10000\, {\rm s^{-1}}$ to $a = 5000\, {\rm s^{-1}}$ on the flamelet) produces an error on consumption speed below 20\% at any filter width, which is still below the maximum error observed for the unstretched manifolds.
Overall, these results indicate that a single strained flamelet can quantitatively reproduce the flame speed in flames subject to differential and preferential diffusion effects and potential subgrid thermodiffusive instabilities, as long as the fixed-strain value is imposed within a reasonable range.
%

Let's now consider the manifolds based on a fixed-strain flamelets with varying equivalence ratio (2DSF in Table~\ref{tab:casesTurb}). The application of varying equivalence ratio in the manifold stems from the fact that turbulent eddies and induced strain and curvature might affect the thermochemical state within the flame.
By looking at the consumption speed ratio in Figure~\ref{fig:ScT}, one can notice that the error now
depends again on the filter width, although more weakly than the 2DU-type manifolds, and its maximum value ranges from 20\% to 30\% for the lower strain case (Fig.~\ref{fig:ScTa2000}), and reduces to below 20\% for the higher strain case (Fig.~\ref{fig:ScTa5000}).
Nevertheless, the error in this case is much less sensitive to the choice of the fixed-strain level imposed on the flamelets. 

\subsubsection{Modelling error} \label{sec:results.turb.modError}
\begin{figure}
\centering
\subfloat[Unfiltered grid. \label{fig:modErrorDNS}]{%
\resizebox*{0.31\textwidth}{!}
{\includegraphics{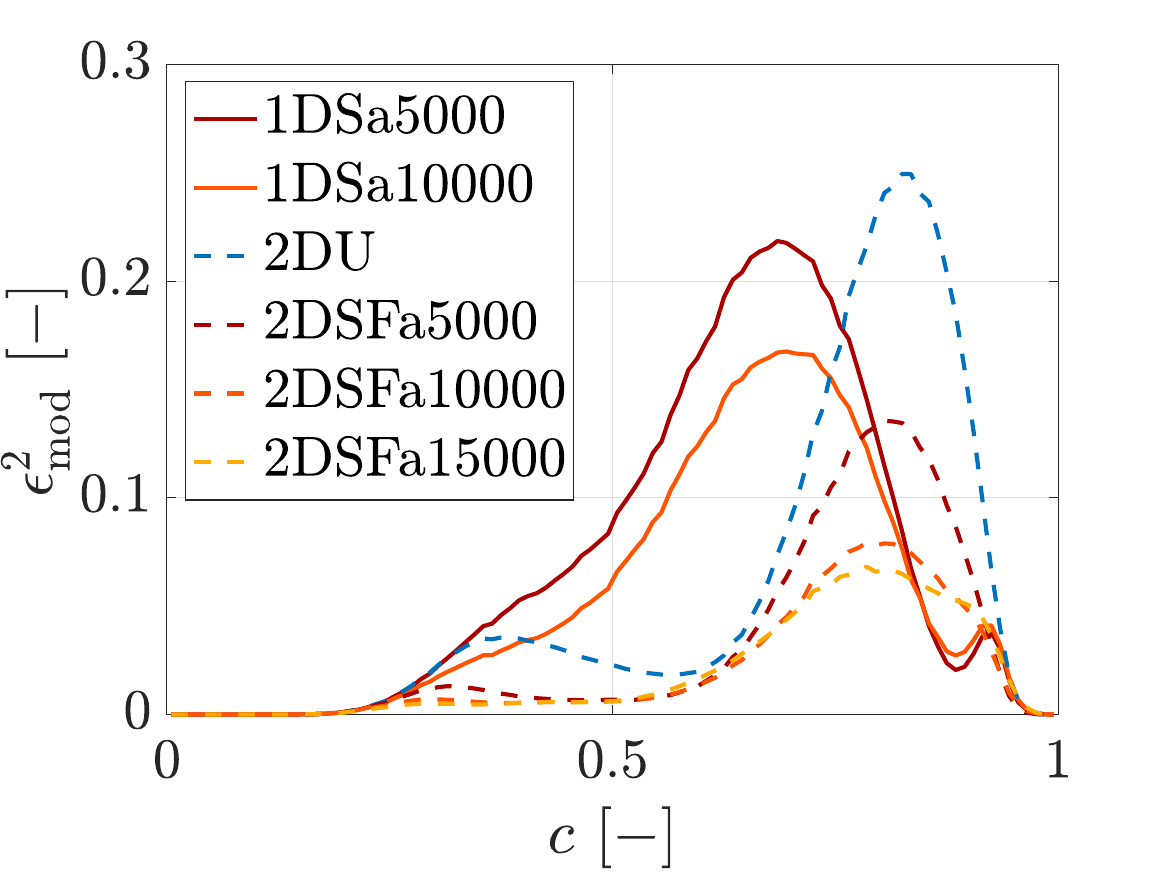}}}\hspace{5pt}
\\
\subfloat[$\Delta=0.5\delta_f$. \label{fig:modError05}]{%
\resizebox*{0.31\textwidth}{!}{\includegraphics{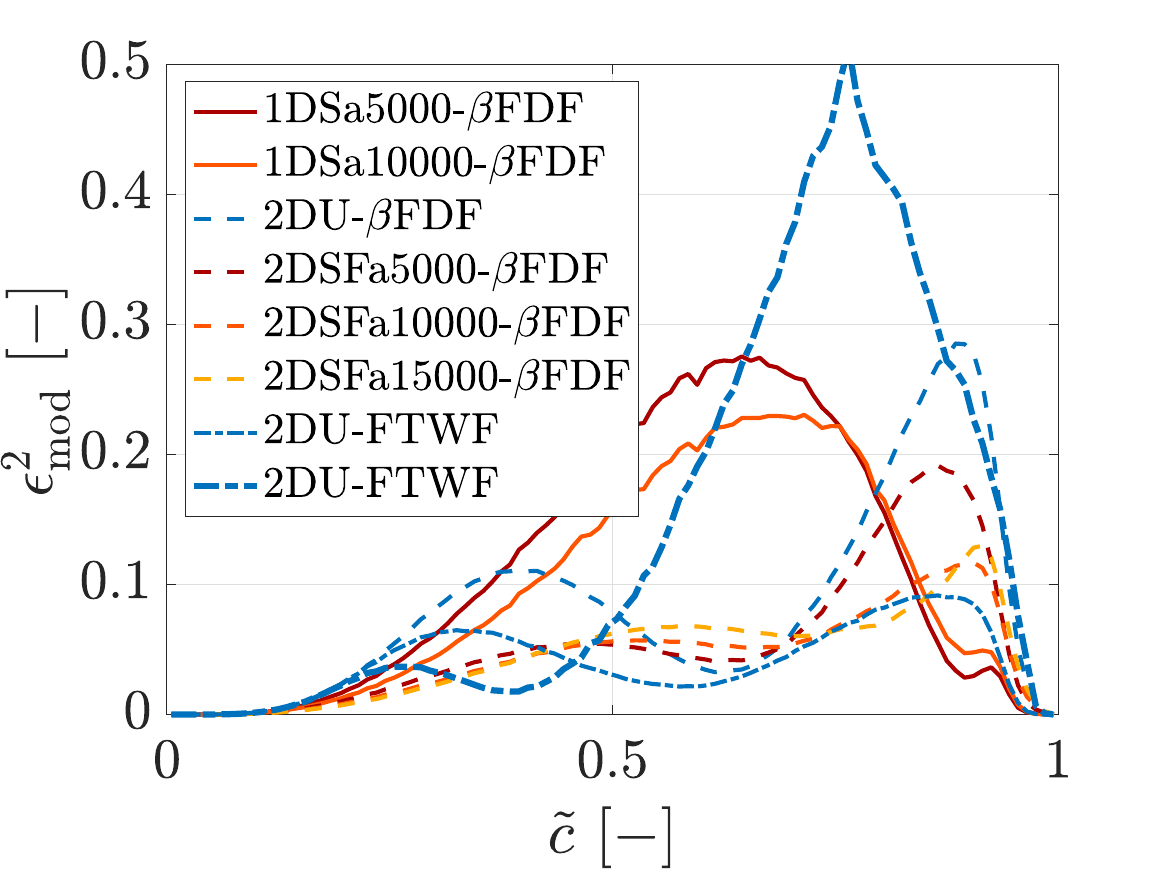}}}\hspace{5pt}
\subfloat[$\Delta=1\delta_f$. \label{fig:modError1}]{%
\resizebox*{0.31\textwidth}{!}{\includegraphics{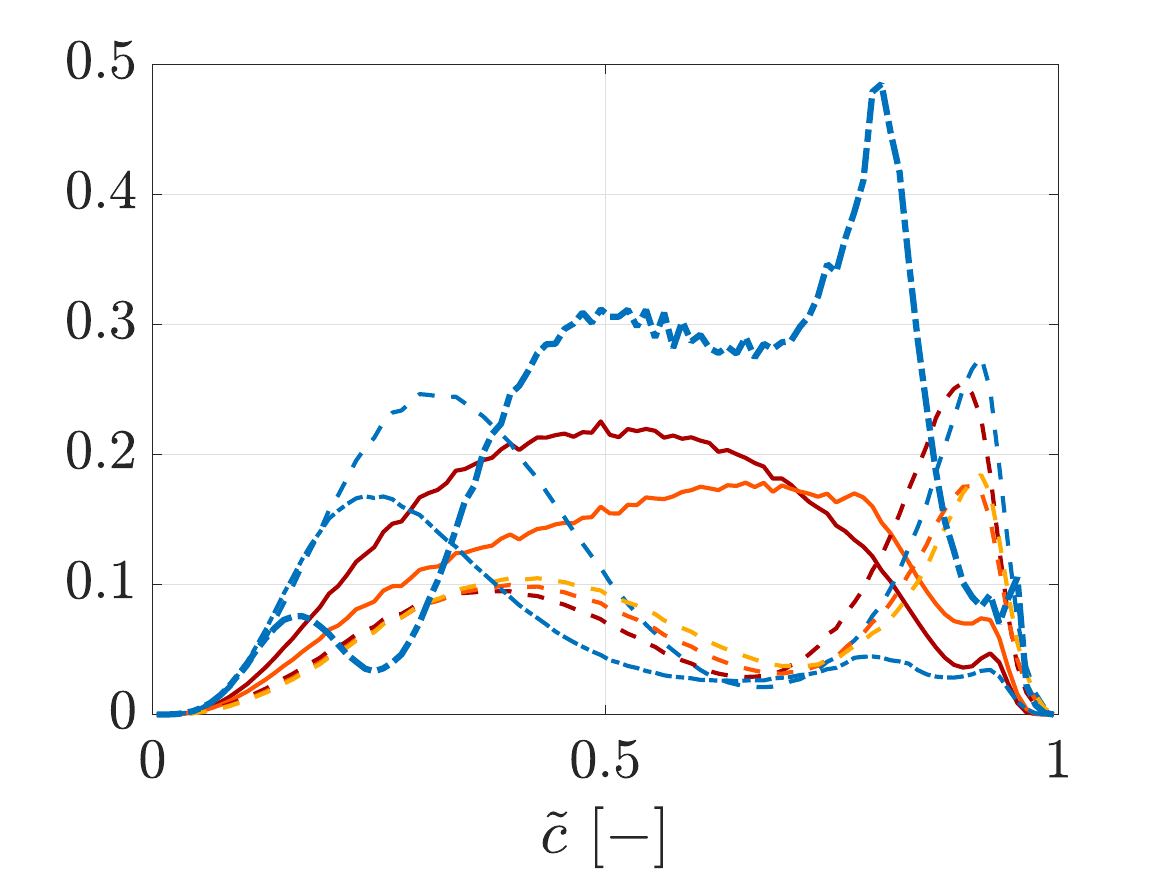}}}\hspace{5pt}
\subfloat[$\Delta=2\delta_f$. \label{fig:modError2}]{%
\resizebox*{0.31\textwidth}{!}{\includegraphics{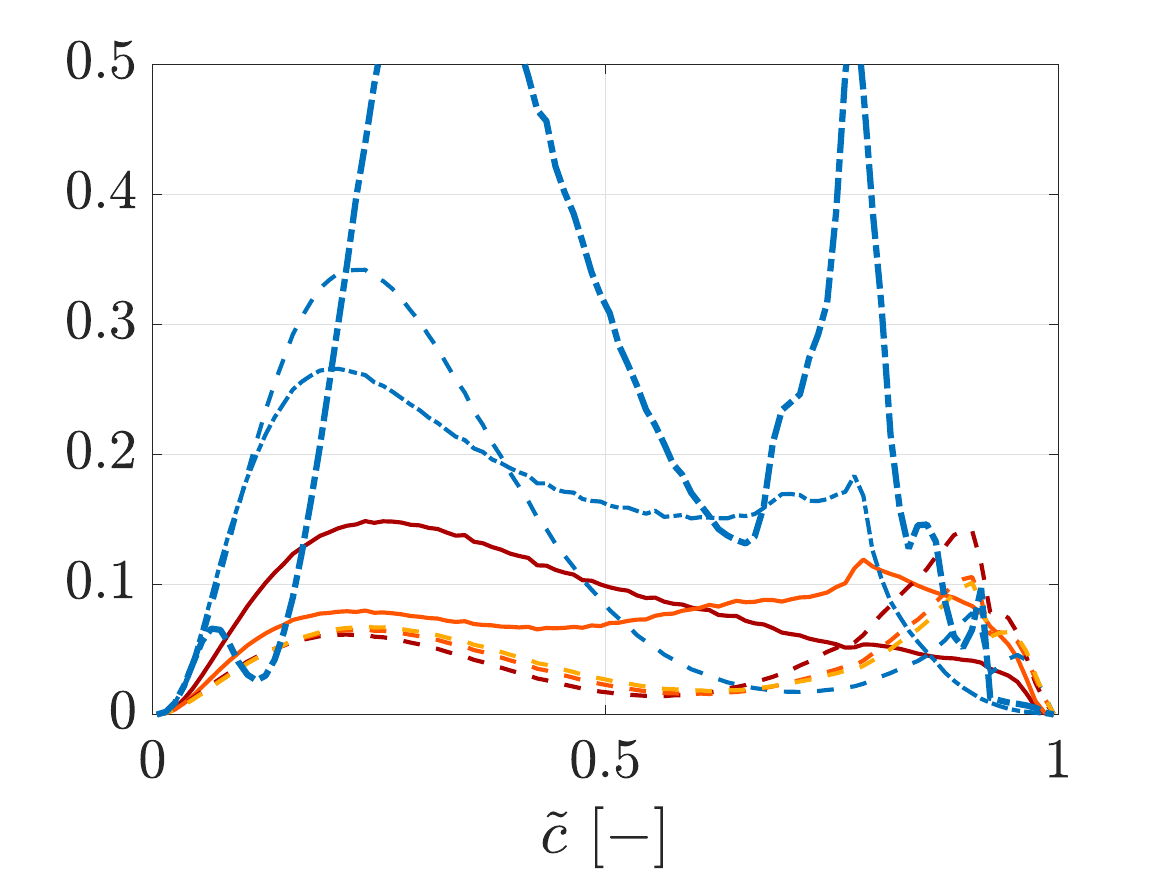}}}
\caption{Conditionally averaged modelling error at a random time $t=t_1$ of the  turbulent DNS flame with applied strain rate $a=5000\, {\rm s^{-1}}$, for unfiltered (top) and filtered (bottom) mesh at different filter widths. The nomenclature in the legend refer to
    Table~\ref{tab:casesTurb}.} \label{fig:modError}
\end{figure}

\begin{figure}
    \centering
    \includegraphics[width=0.9\textwidth]{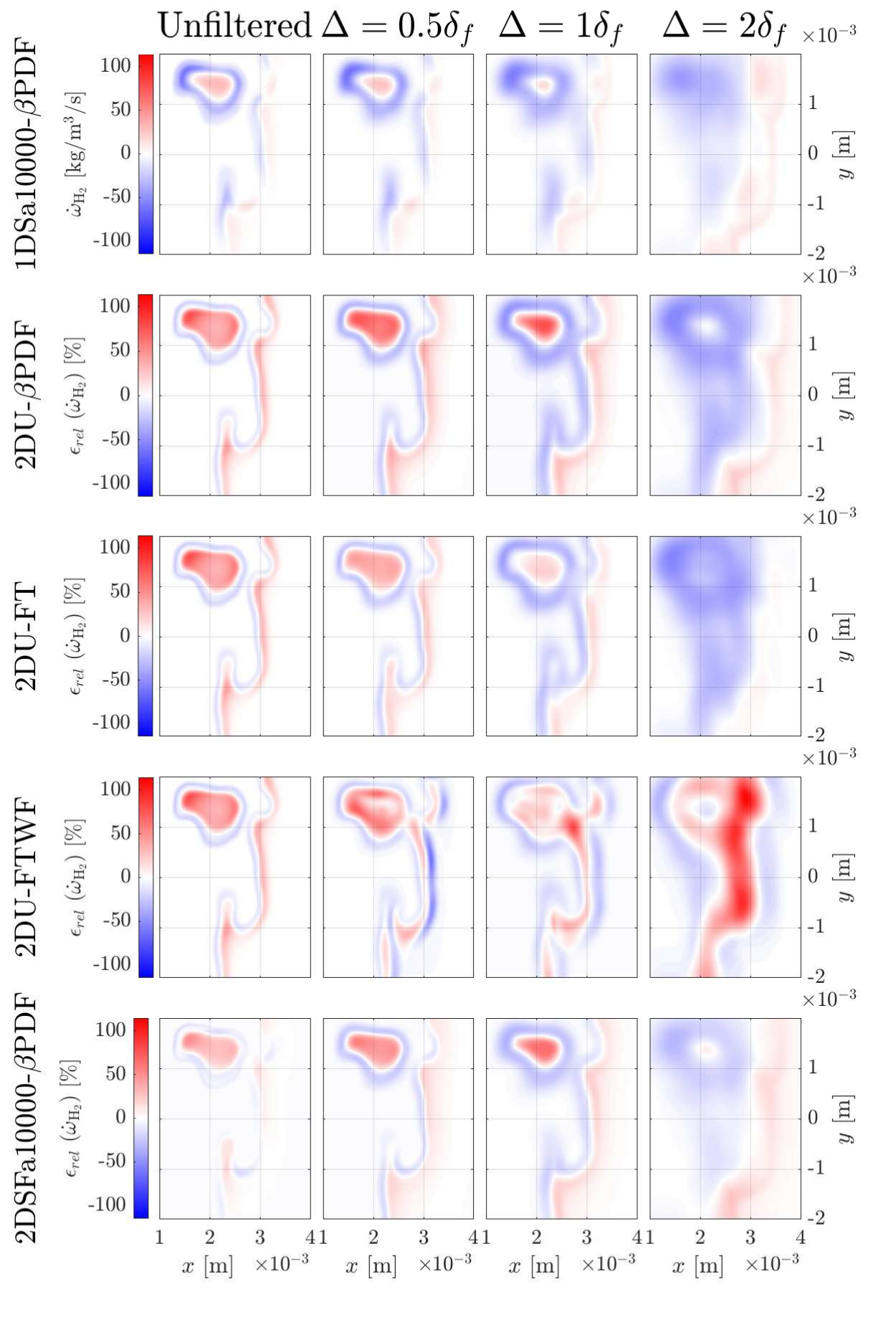}
    \caption{Midplane contours of relative error of the H$_2$ source term $\dot{\omega}_{\rm H_2}$ (Eq.~\eqref{eq:epsRel}) for increasing filter widths and some of the manifold approaches of Table~\ref{tab:casesTurb}.}
    \label{fig:a5000TCont}
\end{figure}

Additional insight on the performance of the various manifold approaches of Table~\ref{tab:casesTurb} is provided by quantifying the conditionally averaged modelling error for the fuel reaction rate, which is defined as~\cite{berger2025combustion}:
\begin{equation}
    \label{eq:modError}
    \epsilon^2_{mod} = \frac{ \left \langle (\dot{\omega}_{\rm H_2}^M -  \dot{\omega}_{\rm H_2}^{\rm ref} ) ^2 |c \right \rangle}{\rm max (\langle \dot{\omega}_{\rm H_2}^{\rm ref} | c \rangle )^2}.
\end{equation}
This definition allows to assess how the modelling error varies across the flame front as compared to the peak mean reaction rate. The error variation in filtered progress variable space is shown in Figs.~\ref{fig:modErrorDNS} and~\ref{fig:modError05}-\ref{fig:modError2} respectively for the unfiltered and filtered (at different $\Delta$) meshes.
Midplane contour plots of the relative error according to Eq.~\eqref{eq:epsRel} are further shown 
in Figure~\ref{fig:a5000TCont} for a subset of cases in Table~\ref{tab:casesTurb}.
%
Only the DNS case with applied strain rate $a=5000\, {\rm s^{-1}}$ is shown as similar observations could be made for the case at $a=2000\, {\rm s^{-1}}$.

Consistently with the laminar flames analysis (Section~\ref{sec:results.lam}) and the turbulent flame consumption speed assessment, the manifold of unstretched flamelets exhibits the highest modelling error on the unfiltered grid (see Fig.~\ref{fig:modErrorDNS}), with an average deviation from the DNS reaction rate of about 25\%. Accordingly, the highest relative errors are observed in space in the contour plots of Figure~\ref{fig:a5000TCont}. 
The modelling error seems to only mildly reduce when the single fixed-strain flamelet is employed (even with the imposed strain rate of $a=10000\, {\rm s^{-1}}$ giving the smallest error on consumption speed).
The error seems to reduce more significantly, instead, when a manifold of fixed-strain flamelets with varying equivalence ratio is used. The error peak in this case (see $\widetilde{c}\approx 0.7$ in Fig.~\ref{fig:modError}) reduces further when increasing the applied strain rate in the flamelets from $a=5000\, {\rm s^{-1}}$ to $a=10000\, {\rm s^{-1}}$, but shows negligible additional improvement for $a=15000\, {\rm s^{-1}}$, the reasons for which are discussed in the next subsection.

When the mesh is filtered, the modelling error for the unstretched flamelets manifold and presumed $\beta$-FDF reaches values of about 30\% at all filter widths. Furthermore, two peaks are observed, respectively at low values of progress variable (corresponding to underestimation of reaction rate, see Sec.~\ref{sec:results.lam.2DU}) and high values of progress variable (corresponding to overestimation of reaction rate). In detail, the reaction rate is overall more overestimated at small filter sizes ($\Delta=0.5\delta_f$, the error peak is stronger for large values of progress variable), and underestimated at large filter sizes (see $\Delta=2\delta_f$, where the error peaks is stronger at low values of progress variable) confirming the strong filter-dependence of this manifold approach. At intermediate filter widths ($\Delta=\delta_f$), the two peaks are observed to be of the same magnitude, suggesting that the errors might compensate each other in this case, leading to about correct consumption speed observed for Figure~\ref{fig:ScT}. 

When the F-TACLES approach is used without wrinkling factor (with the unstretched flamelets manifold), the modelling error is observed to remain below 10\% for small filter sizes ($\Delta=0.5\delta_f$), but the peak for low values of progress variable (underestimation of reaction rate) significantly increases to values above 20\% for larger filter sizes. Unlike for the observation on the consumption speed in Section~\ref{sec:results.turb.sc}, however, introducing the wrinkling factor does not reduce the modelling error of Eq.~\eqref{eq:modError}, and instead results in sharp peaks of about 50\%. 
Consistently, the contour plots of Figure~\ref{fig:a5000TCont},
show that for increasing filter sizes the relative error increases in magnitude in the F-TACLES approach when the wrinkling factor is introduced. However, this errors corresponds to both regions of overestimation and underestimation of reaction rate, which apparently compensate for each other when integrating the reaction rate to compute the consumption speed. This suggests that the wrinkling factor would introduce limitations in mimicking the correct local behavior in terms of differential and preferential diffusion effects.

Let's now consider the performance of fixed-strain single-flamelet manifold with presumed $\beta$-FDF at increasing filter widths.
Here the modelling error exhibits a similar behaviour to that observed for the irreducible error examined in Section~\ref{sec:results.turb.irrErr}. The error peak in this case is relatively high, between 20\% and 30\%, for the smallest filer width ($\Delta=0.5\delta_f$), but decreases to values around 10\% for $\Delta=2\delta_f$. Note that the error is lower when a strain rate of $a=10000 \, {\rm s^{-1}}$ is applied on the flamelet since this value is closer to the value experienced by the turbulent flame. The higher error at smaller filter sizes is explained by the fact that a single flamelet is not able alone to capture the correct thermochemical states in correspondence of the local leaning and enrichment of mixture fraction induced by preferential and differential diffusion effects in strained and curved regions of the flame, corresponding to modified local reactivity. The fact that the error in the consumption speed is still very low for this manifold is therefore probably due to compensating overestimation and underestimation of reaction rates in correspondence of leaner and richer regions, respectively. The improvement at larger filter sizes is due to the fact that these local phenomena are moved more and more to subgrid scales at increased filter widths, allowing to observe at the resolved scales only the related overall increased flame reactivity induced synergistically by thermodiffusive instabilities and turbulence~\cite[see][for example]{berger2022synergistic}. As proved by this analysis, this globally increased flame reactivity can be captured with tabulated chemistry models by simply tuning the applied strain rate of a single flamelet.

Lastly, the manifold with fixed-strain flamelets of varying equivalence ratio is assessed. This manifold approach leads to the smallest modelling error peaks among the cases of Table~\ref{tab:casesTurb}, which are observed in Figure~\ref{fig:modErrorDNS} to be overall contained around 10\% or below. Similar considerations can be driven by inspecting the contour plot of the relative error for this manifold in Figure~\ref{fig:a5000TCont}. 
This suggests that this approach, unlike the others, has potential to mimic an about correct local behaviour in terms of differential and preferential diffusion effects in strained and curved regions. The fact that other manifold approaches lead to reduced errors on the consumption speed at specific meshes is thus to be interpreted as compensation of errors in the parametrisation as demonstrated by the present analysis (a part from the 1DS-type manifold at large filters discussed in the previous paragraph). A stronger sensitivity to the choice of the fixed-strain value in the manifold (which is generally not known \textit{a priori} in a LES context) is observed as compared to the results for the consumption speed. This dependence is further investigated in the next section.

\subsection{Choice of fixed strain rate value} \label{sec:results.manifoldStrain}
\begin{figure}
    \centering
    \includegraphics[width=0.5\textwidth]{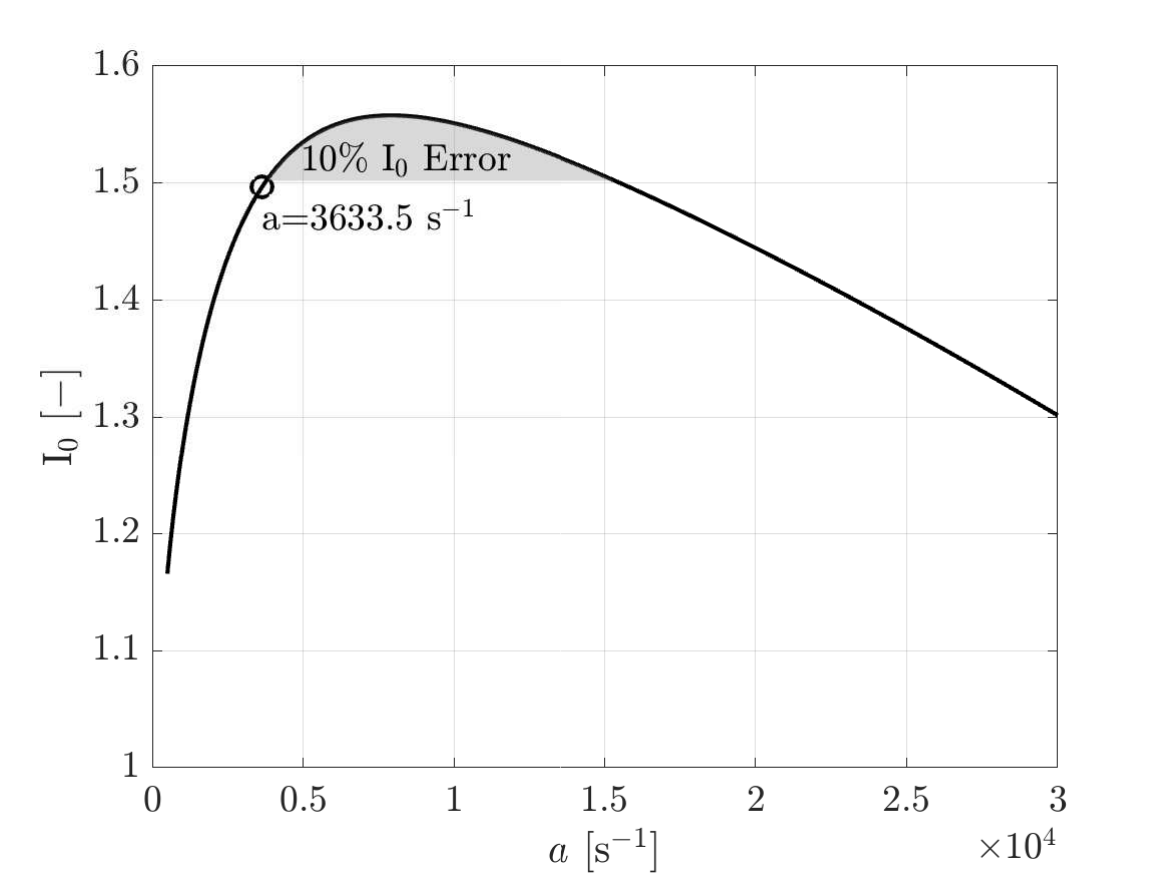}
    \caption{Stretch factor of a laminar flamelet as a function of the applied strain rate.}
    \label{fig:I0a}
\end{figure}

Although improved performance was observed in the \textit{a priori} analyses of fixed-strain flamelets manifolds for both laminar and turbulent conditions (Tables~\ref{tab:cases} and~\ref{tab:casesTurb} respectively), the values of strain imposed on the flamelets in these analyses could be chosen by inspection of the DNS data, which is generally not possible in the context of a LES.
An analysis is thus conducted here to assess how the choice of the fixed-strain value influences the reconstructed reaction rate. For this purpose the stretch factor ${\rm I}_0 = S_c/s_L$ of a laminar flamelet at the nominal equivalence ratio $\phi=0.5$ across different strain rates is computed, and presented in Figure~\ref{fig:I0a}. As observed, the stretch factor exhibits a peak value of approximately 1.55 in the strain rate range of 7000 s$^{-1}$ to 8000 s$^{-1}$, indicating a 55\% increase in consumption speed as compared to the laminar unstretched flame speed. 
Notably, the stretch factor shows a steep increase with strain rate up to values of about $a=3600 \, {\rm s^{-1}}$, followed by a more gradual variation in the region near the peak.
For values of strain rate ranging between about 3600 s$^{-1}$ to 15000 s$^{-1}$ (shaded area in the figure), a misprediction of applied strain rate would lead to a maximum 10\% error in the stretch factor.
%
This range includes the strain rates of all tested manifolds of strained flamelets in the turbulent setting. This result suggests that imposing any strain rate on the flamelet within this range would lead to similar predictions, at least for integral quantities like the consumption speed, which is consistent with the observations in Figure~\ref{fig:ScT}.

This outcomes suggests that, at least for the ranges of turbulence and strain investigated in this study, relatively low-dimensional manifolds consisting of a single strained flamelet can be used in a LES with relatively large filter sizes to predict, with limited errors, integral quantities such as consumption speed, even in presence of preferential and differential diffusion effects. 
A set of flamelets with varying equivalence ratio, but still with fixed value of strain, might be accurate enough to further predict local fluctuations of reaction rates induced by preferential and differential diffusion effects, such as strain and curvature effects and the onset of thermodiffusive instabilities. This represents an important simplification for flamelets-like models as strained flamelets with varying levels of strain, implying a higher-dimensional manifold, might be avoided. However, this should be ultimately assessed \textit{a posteriori} in future works.

\subsection{Unstretched flamelets correction} \label{sec:results.correction}
Previous sections have demonstrated both poor \textit{a priori} performance and strong filter dependence in models based on manifolds constructed from unstretched flamelets. Specifically, unlike strained flamelets, unstretched flamelets provided a poor prediction of the consumption speed already in laminar conditions, showing a counter-intuitive systematic decrease with increasing filter width. This systematic decrease was more significant at higher simulation strain rates, suggesting that the reaction rate map it provides is not a good picture of the thermochemical states of lean premixed and turbulent hydrogen flames, as later proved in Section~\ref{sec:results.turb}. In light of this, here we propose a correction for the predicted turbulent consumption speed based on the results of the laminar analysis. 

At the highest tested strain rate, the predictions in the laminar analysis from both subgrid closures exhibited perfect agreement (see Fig.~\ref{fig:L2DU}). This level of strain is marked in Figure~\ref{fig:I0a}, showing that its corresponding stretch factor belongs to the 10\% band from the peak. A correction function $f(\Delta/\delta_f)$, valid for the range of strain rate considered, can therefore be derived from these laminar data, such that
\begin{equation}
    \label{eq:correction}
    \frac{S_c^*}{S_{c,\rm DNS}} = f\left(\frac{\Delta}{\delta_f}\right) \frac{S_c^{\rm 2DU}}{S_{c,\rm DNS}},
\end{equation}
where $S_c^{\rm 2DU}$ represents the consumption speed reconstructed using the 2DU manifold, and $S_c^*$ denotes the corrected consumption speed. The correction function is defined as the inverse of $\frac{S_c^{\rm 2DU,lam}}{S_{c,\rm DNS}^{\rm ref,lam}}\left(\frac{\Delta}{\delta_f}\right)$, obtained from the 2DU manifold applied to laminar simulations at $a=3633.5$ s$^{-1}$. To determine this function, we perform a double exponential fit using combined data points from both the $\beta$-FDF and F-TACLES models across all tested filtered grids. Note that the 2DU-FTWF tabulation with the wrinkling factor has not been corrected here due to the limitations highlighted in Section~\ref{sec:results.turb.modError}.

\begin{figure}
\centering
\subfloat[Case a2000T. \label{fig:a2000T2DUCorr}]{%
\resizebox*{0.48\textwidth}{!}{\includegraphics{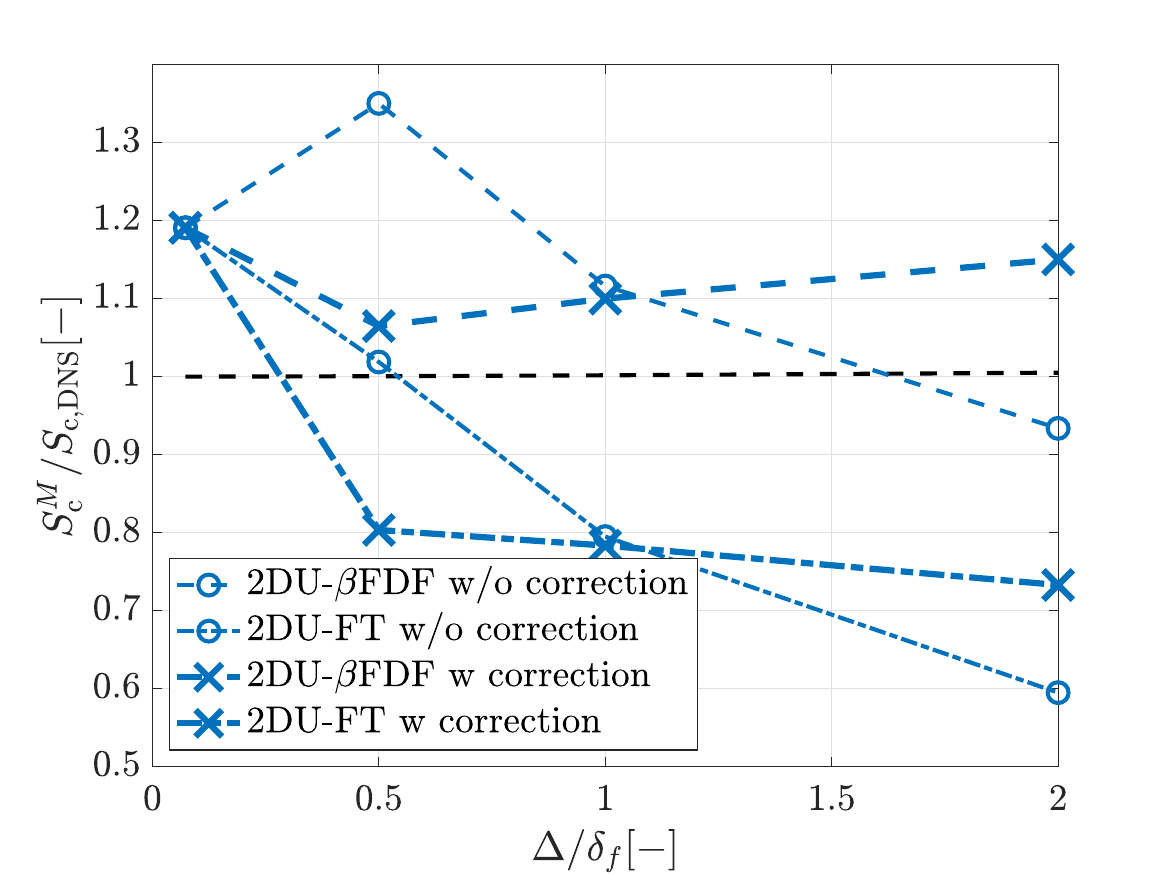}}}\hspace{5pt}
\subfloat[Case a5000T. \label{fig:a5000T2DUCorr}]{%
\resizebox*{0.48\textwidth}{!}{\includegraphics{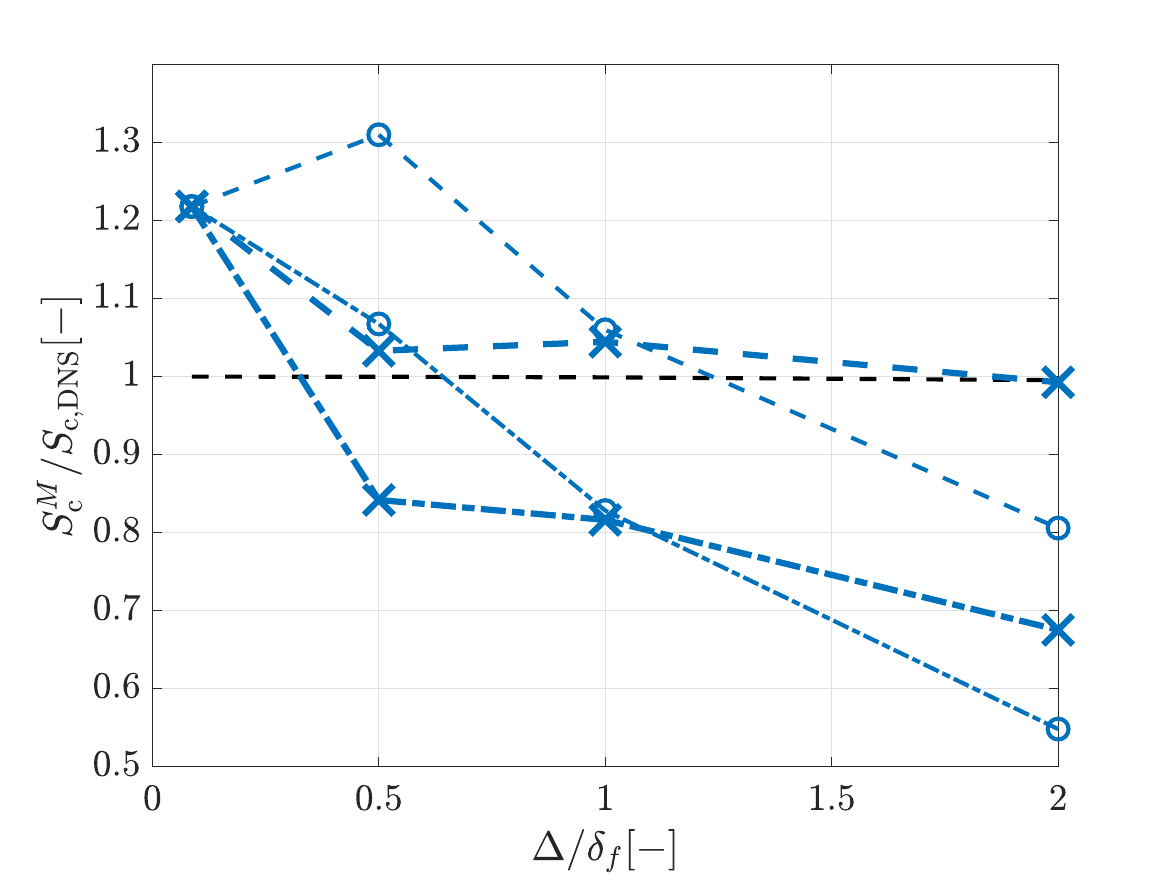}}}
\caption{A priori consumption speeds for the turbulent cases obtained with 2DU type manifolds, with and without the proposed correction with increasing filter width. The reader is referred to Table~\ref{tab:casesTurb} to interpret the graph legend. For a single snapshot of case a2000T (a) and a5000T (b).} \label{fig:2DUCorr}
\end{figure}

This correction is applied to both the 2DU-$\beta$FDF and 2DUFT tabulation predictions, with the results shown in Figures~\ref{fig:a2000T2DUCorr} and~\ref{fig:a5000T2DUCorr} for the cases a2000T and a5000T, respectively.
The figures show that the correction significantly improves the 2DU-$\beta$FDF tabulation predictions. In both the a2000T and a5000T cases, the corrected consumption speed exhibits markedly reduced filter dependence, thereby addressing the primary limitation of this tabulation identified in previous sections. Moreover, the error decreases to below 15\% in the lower strain rate case and to below 5\% in the higher strain rate case. These results demonstrate that the weaknesses observed in 2DU manifold predictions stem solely from intrinsic tabulation errors rather than inaccuracies in the way the $\beta$-FDF subgrid model captures turbulence-chemistry interactions. Although these inaccuracies originate from incorrectly reconstructed local reaction rates, they can be effectively corrected at the integral level, corresponding to the grid cell scale in finite-volume CFD solvers, with a methodology similar to that suggested by Nilsson \textit{et al}.~\cite{nilsson2019priori}---through a correction derived from laminar simulations at strain rates near the peak stretch factor. This correction will be tested \textit{a posteriori} in a future work.

Considering the 2DU-FT tabulation, on the other hand, the correction is not as effective. Although there is a moderate improvement, the prediction of the consumption speed remains substantially dependent on the width of the filter, and an underestimate above 20\% remains. This proves that the proposed correction is not able alone to correct the reaction rate to account for subgrid wrinkling, but only for the intrinsic tabulation error of manifolds made of unstretched flamelets. Indeed, to further improve the predictions, this correction should be combined with an appropriate wrinkling factor; however, such an approach has not been pursued in the current work due to challenges encountered in implementing the wrinkling factor correction, as discussed in Section~\ref{sec:results.turb.modError}.

\section{Conclusions} \label{sec:conclusions}
This study presents a comprehensive \textit{a priori} assessment of various tabulated-chemistry models for lean premixed hydrogen combustion with differential and preferential diffusion, with particular focus on addressing the challenges posed by the synergistic effect of thermodiffusive instabilities and turbulence in strained counterflow settings. The main findings obtained through systematic analysis of both laminar and turbulent flame configurations are summarised below.
\begin{itemize}
    \item Limitations of Unstretched Flamelet Manifolds: The investigation highlights significant inaccuracies in the thermochemical states parametrisation based on unstretched flamelet manifolds.
    These manifolds exhibit systematic overestimation of reaction rates at high progress variables and underestimation at low progress variables.
    While in some cases correct predictions are obtained by compensation of these effects, the consumption speed obtained through this manifold are overall unreliable, and worsen with increasing strain rates and filter widths, thus requiring and \textit{ad hoc} correction.
    \item Effectiveness of Strained Flamelet Approaches: Fixed-strain flamelet manifolds demonstrate improved performance compared to unstretched counterparts. When the flamelet strain rate matches the total strain rate of the simulated flame, the single strained flamelet manifold shows a remarkably low modelling error at large filter sizes. The novel strained flamelets manifold, combining fixed strain with varying equivalence ratio, offers enhanced ability to reconstruct the local reaction rates determined synergistically by thermodiffusive instabilities and turbulence at all tested filter sizes.
    \item Strain Rate Selection Guidelines: Analysis of the stretch factor reveals a wide range of strain rates (3500-15000 s$^{-1}$) where the corresponding stretch factor changes by less than 10\%. The strain rate for fixed-strain flamelets manifolds should be selected within this range to ensure the model predictions remain relatively consistent in a LES framework, where the simulation strain rate is unknown a priori.
    \item Correction Methodology for Unstretched Manifolds: A novel correction function derived from laminar simulations significantly improves the predictions \textit{a priori} of integral quantities such as the consumption speed of unstretched flamelet manifolds, reducing filter dependence. While effective for $\beta$-FDF models, this correction proves less successful for F-TACLES implementations, suggesting the need for combined approaches with wrinkling factors.
\end{itemize}

Overall, for coarse grid simulations, ultra-light manifolds consisting of single strained counterflow flamelets with appropriate strain rates and $\beta$-FDF subgrid closures offer computationally efficient and reliable solutions to capture the increased reactivity at subgrid scales determined by thermodiffusive instabilities, turbulence, and strain. However, it's important to note that this approach works effectively only at coarser grids, as finer resolutions would reveal the model's inability to capture local mixture fraction and reaction rate oscillations, potentially leading to poor \textit{a posteriori} performance. For higher fidelity requirements across multiple grid resolutions, both the corrected unstretched flamelets manifold and the novel manifold with fixed strain and varying equivalence ratio with $\beta$-FDF closure provides reliable predictions of integral quantities such as the consumption speed. Ultimately, the novel strained manifold shows the best performance among all the tabulations tested in reconstructing the local reaction rates.

Unlike previous works, this study achieves modelling improvements without increasing the dimensionality of the manifolds, thereby maintaining computational efficiency and keeping memory costs unchanged. 
Future work should focus on \textit{a posteriori} validation of these findings and further refinement of correction methodologies for practical LES applications.

\section*{Acknowledgements}

A.P. and I.L. acknowledge the Dutch Ministry of Education and Science for providing funding support to this project via the Sector Plan scheme. A.P further gratefully acknowledges financial support to perform Short Term Scientific Mission (STSM) at Sapienza University (Rome) from the CYPHER consortium funded by the European Cooperation in Science and Technology (COST ACTION CA22151). I.L. further gratefully acknowledges financial support from the ERC Starting Grant OTHERWISE, grant n. 101078821. P.E.L. and F.C. acknowledge financial support by ICSC (Centro Nazionale di Ricerca in HPC, Big Data and Quantum Computing) funded by the European Union – NextGenerationEU. 

\section*{Disclosure statement}

The authors declare that there is no conflict of interest.








\bibliographystyle{unsrtnat}
\bibliography{references}




\end{document}